\documentclass[prd,aps,twocolumn,floats,showpacs]{revtex4}

\usepackage{graphicx}
\usepackage{dcolumn}
\usepackage{amsmath}
\usepackage{epsf}

\def\maxima{{\sc maxima}}
\def\maximai{{\sc maxima-i}}

\def\simlt{\lower.5ex\hbox{$\; \buildrel < \over \sim \;$}}
\def\simgt{\lower.5ex\hbox{$\; \buildrel > \over \sim \;$}}

\def\eg{{\rm e.g.}}
\def\ie{{\rm i.e.}}
\def\cf{{\rm cf.}}
\def\etal{{\rm et al.}}
\def\eqn{{\rm Eq.}}
\def\eqns{{\rm Eqs.}}

\def\order#1{${\cal O}\left(#1\right)$}
\def\bm#1{{\hbox{\boldmath $#1$\unboldmath}}}

\def\myatop#1#2{\begin{array}{c} \displaystyle{#1} \\ \displaystyle{#2}\end{array}}
\def\dsp{\displaystyle}
\def\l#1{\left#1}
\def\r#1{\right#1}

\def\apj{{\sl Ap.J.}}
\def\apjs{{\sl Ap.J.Supp.}}
\def\apjlett{{\sl Ap.J.}}
\def\aa{{\sl A\&A}}
\def\mnras{{\sl MNRAS}}

\def\prd{{\sl Phys.\ Rev.\ D.}}

\def\wtilde{\widetilde}  

\def\gA{{\cal A}}
\def\gN{{\cal N}}
\def\gm{\bm{\hat m_p}}
\def\iNtt{\bm{N_t}^{-1}}
\def\iNct{\bm{N_{Ct}}^{-1}}

\def\htaum{{\bm{\tilde \tau_{\l(m\r)}}}}

\begin{document}

%\baselineskip 1.5cm

%\twocolumn[\hsize\textwidth\columnwidth\hsize\csname
%@twocolumnfalse\endcsname
%%
%%
%\tighten

\title{Making Maps Of The Cosmic Microwave Background:\\ The {\sc maxima} Example.}

\author{
 Radek Stompor$^{1,2,3}$,
 Amedeo Balbi$^{4}$, 
 Julian D.\ Borrill$^{5,1}$,
 Pedro G.\ Ferreira$^{6}$,
 Shaul Hanany$^{7,1}$,
 Andrew H.\ Jaffe$^{1,8,9}$,
 Adrian T.\ Lee$^{9,1,10}$,
 Sang Oh$^{1,9}$,
 Bahman Rabii$^{1,9}$,
 Paul L.\ Richards$^{1,9}$,
 George F.\ Smoot$^{1,10,2}$,
 Celeste D.\ Winant$^{1,9}$,
 Jiun-Huei Proty Wu$^{8}$
}
\affiliation{
$^1$ Center for Particle Astrophysics, University of California, Berkeley, CA 94720, USA\\
$^2$ Space Sciences Laboratory, University of California, Berkeley, CA 94720, USA\\
$^3$ Copernicus Astronomical Center, Warszawa, Poland\\
$^4$ Dipartimento di Fisica, Universit{\`a} Tor Vergata, Roma, Italy\\
$^5$ NERSC, Lawrence Berkeley National Laboratory, Berkeley, CA 94720, USA\\
$^6$ Astrophysics, University of Oxford, NAPL, Oxford, OX1 3RH, UK\\
$^7$ School of Physics and Astronomy, University of Minnesota, Minneapolis, MN55455, USA\\
$^8$ Dept. of Astronomy, University of California, Berkeley, CA 94720, USA\\
$^{9}$ Dept. of Physics, University of California, Berkeley, CA 94720, USA\\
$^{10}$ Lawrence Berkeley National Laboratory, Berkeley, CA 94720, USA
}
\begin{abstract}
\vskip 0.5truecm
This work describes Cosmic Microwave Background (CMB) data analysis
algorithms and their implementations, developed to produce a pixelized map
of the sky and a corresponding pixel-pixel noise correlation matrix
from time ordered data for a CMB mapping experiment.
We discuss in turn algorithms for estimating noise properties from the
time ordered data, techniques for manipulating the time ordered data,
and a number of variants of the maximum likelihood map-making
procedure. We pay particular attention to issues pertinent to {\em
real} CMB data, and present ways of incorporating them within the
framework of maximum likelihood map-making. Making a map of the sky is
shown to be not only an intermediate step rendering an image of
the sky, but also an important diagnostic stage, when tests for
and/or removal of systematic effects can efficiently be performed.
The case under study is the \maximai\ data set. However, the
methods discussed are expected to be applicable to the analysis of other
current and forthcoming CMB experiments.
\end{abstract}


% 98.70.Vc - cosmic backgrounds
% 98.80.Bp - origin and formation of the universe
% 98.80.Es - observational cosmolofy

\maketitle

\section{Introduction}

This paper presents a comprehensive set of data analysis methods
aiming at the production of a map of the sky and an accurate estimate
of map uncertainty in a case of CMB mapping experiments.
We describe a variety of maximum-likelihood-based
map-making methods, and discuss their performance in the analysis of the
\maximai\ data set. 

\maxima\ is a balloon-borne experiment~\cite{Lee1999} built
primarily in Berkeley~\cite{Maxima}
and designed to make a number of
short-duration flights. To date the \maxima\ team has published the
results of the first flight of the
instrument~\cite{Hanany2000,Lee2001}, consisting of a high-resolution
map of almost $100$ square degrees of the microwave sky, together with
a power spectrum of the CMB anisotropies observed in the map covering
a broad range in $\ell$ space from $\ell \sim 35$ up to $\sim 1235,$
corresponding to angular scales from $5$ degrees down to $5$
arcminutes. Such products are final results of an involved data
analysis pipeline described in this paper.  
The complexity and size of this data set have proven
to be a significant challenge for data analysis methods setting
demanding requirements for both their precision and speed. The challenge
which our
methods and tools are designed to meet.
With other complex and advanced CMB experiments in progress and anticipated (including
the satellite missions, MAP~\cite{Map} and
Planck~\cite{Planck}), these tools and methods
can be expected to be of wider interest and applicability.
Describing the details of the \maximai\ data analysis is another goal 
of this paper.

The structure of this work is as follows; in Sect.~\ref{timedomain_main:para} we deal with
the data in the time domain, focusing on data pre-processing and noise
estimation, including an outline of the basic features of the
\maximai\ data set, and of the simulation tools used to test our
map-making pipeline. Sect.~III is devoted to the description and
comparison of a suite of different map-making methods. Those 
simultaneously produce both a map and a corresponding pixel-pixel noise correlation matrix.
Although the
algorithms are all based on the maximum likelihood approach, they
differ in the way they attempt to optimize the balance between
accuracy and speed. We demonstrate their performance in analyzing
\maxima-like simulations as well as the actual \maximai\ data set.
In Sect.~IV we discuss ways of handling systematic effects within
the general framework of maximum likelihood map-making. Although such
systematics are inevitable in real CMB data sets, they are rarely
considered in more theoretical accounts of CMB data analysis (\eg,
~\cite{Wright1996a,Tegmark1997maps,Tegmark1997design,Oh1999,
Ferreira2000,Stompor2000,Hinshaw2000,Dore2001a,Natoli2001}).  In Sect.~V we
combine these elements, and consider some practical aspects of
recently-proposed iterative algorithms for time-domain noise
estimation~\cite{Ferreira2000, Prunet2000}. In Sect.~VI, we
complete our presentation with a description of the numerical tests we
have developed to check consistency of our analysis.

\begin{figure}[t]
\vskip 0truecm
\leavevmode\epsfxsize=11.cm\epsfbox{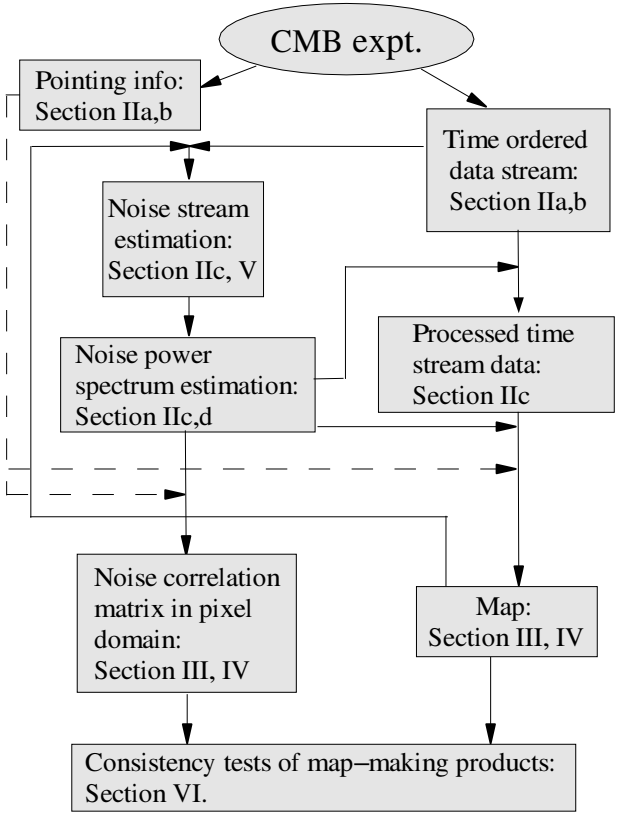}
\vskip -3.5truecm
\caption{A flow-chart showing the layout and inter-dependencies of the
sections of this paper, as well as mutual relations of different stages
of the data analysis pipeline as described in the text.}
\label{flowchart:figure}
\end{figure}

The inter-dependencies of different sections of this paper are
depicted in Fig.~\ref{flowchart:figure}.

\begin{table}%[t]
\caption{A summary of the notation used in this paper.
\label{notation:table}}
\begin{ruledtabular}
\begin{tabular*}{\hsize}{l@{\extracolsep{0ptplus1fil}}c@{\extracolsep{0ptplus1fil}}c}
%\begin{tabular}{l c c}
Symbol & Description & Section\\
\colrule
$\Delta$  & data sampling interval & \ref{maxdata:para}\\
$\bm{d}(t)$ & time stream data & \ref{timestream:para}\\
$\bm{d_F}(t)$ & time stream data convolved with filter $\bm{F}$& \ref{timestream:para}\\
$\bm{F}$&instrumental filters (all together) &  \ref{timestream:para}\\
$\bm{F_{low}}$&AC low-pass electronic filter &  \ref{timestream:para}\\
$\bm{F_{high}}$&AC high-pass electronic filter &  \ref{timestream:para}\\
$\bm{F_{bolo}}$&bolometer low-pass filter&  \ref{timestream:para}\\
$\bm{t}_{sky}$ & sky signal &   \ref{timestream:para}\\
$\bm{x}\l(\bm{\alpha}\r)$ & $\alpha$-synchronous signal &  \ref{timestream:para}\\
$\bm{n_t}$ & time stream noise & \ref{timestream:para} \\
$P\l(f\r)$ & time domain noise power spectrum &  \ref{timestream:para}\\
$P_W\l(f\r)$ & prewhitened time domain noise spectrum & \ref{noisestim:para}\\
$\bm{W}$     & prewhitening filter & \ref{noisestim:para}\\
$\lambda_c$  & noise correlation length in time domain & \ref{noisestim:para}\\
$S\l(f\r)$ & spectrum smoothing window function & \ref{timecorr:para}\\
$\bm{N_t}$   & time domain noise correlation matrix & \ref{timecorr:para}\\
$\bm{N_{Ct}}$ & circulant part of $\bm{N_t}$ & \ref{timecorr:para}\\
$\bm{N_{St}}$ & sparse part of $\bm{N_t}$ & \ref{circmaps:para}\\
$\bm{A}$   & pointing matrix of the experiment & \ref{mapmaking:para}\\
$\bm{m_p}$ & pixelized sky $\equiv$ map & \ref{mapmaking:para} \\
$\bm{N_p}$ & noise correlation matrix in pixel domain & \ref{mapmaking:para} \\
$\bm{N_{Cp}}$ & circulant part of $\bm{N_p}$ & \ref{circmaps:para} \\
$\bm{N_{Sp}}$ & sparse part of $\bm{N_p}$  & \ref{circmaps:para} \\
$n_{pix}$  & number of pixels in a map $\bm{m_p}$ & \ref{mapmaking:para}\\
$n_s$      & length of the time stream segment & \ref{mapmaking:para}\\
$\bm{B}, {\cal B}$   & pointing matrices of synchronous effects & \ref{extrapixels:para}\\
$\bm{x_q}$ & extra fictitious pixels & \ref{extrapixels:para}\\
$\gA$      & generalized pointing matrix & \ref{extrapixels:para} \\
$\gm$      & generalized map & \ref{extrapixels:para}\\
${\cal N}$ & generalized noise correlation matrix & \ref{extrapixels:para}\\
$\bm{\tau_t}$ & time domain template & \ref{alternative:para}\\
$\delta^K$    & Kronecker delta & \ref{alternative:para}\\
$\bm{u_p}$    & vector of ones & \ref{singularities:para}\\
$\bm{v_p}$      & singular pixel domain eigen-vector & \ref{singularities:para}, \ref{combining_segments:para}\\
$C_\ell$      & CMB anisotropy power spectrum      & \ref{singularities:para}, \ref{cleaning:para}\\
$\bm{M_p}$    & total signal+noise correlation matrix & \ref{singularities:para}, \ref{cleaning:para}\\
$\bm{S_p}\l(C_\ell\r)$ & CMB signal correlation matrix & \ref{singularities:para}, \ref{cleaning:para}\\
$\bm{N_p}^{-{1/2}}$ & Cholesky factor of matrix \bm{N_p} & \ref{assesment:para}, \ref{constest:para}\\
$\bm{w_p}$ & decorrelated map & \ref{constest:para} \\
\end{tabular*}
\end{ruledtabular}
\end{table}

In this paper we do not consider issues related to the subsequent
statistical investigation of these maps, such as tests for 
Gaussianity or power spectrum estimation.
Our
map-making methods are intended to be as general as possible, and
because they provide both a map and a pixel-domain noise correlation matrix,
they do not restrict the subsequent choice of statistical
tool. Methods for obtaining an angular power spectrum from, or for
searching for non-Gaussianity in, such maps have been described in a
number of recent papers (\eg, \cite{Tegmark1997pow, BJK1998,
Borrill1999pow, Szapudi2000, pseudcl, Dore2001b, Hivon2001}
and~\cite{FMG98, Wu2000, Wu2001b, Cayon2001}, respectively).
The power spectra shown in this paper have been computed using a generic
uncustomized version of a quadratic estimator as implemented in the 
publicly-available MADCAP package~\cite{Madcap,Borrill1999mad}.

A comprehensive discussion of the maps and power spectra produced by
the \maximai\ experiment can be found elsewhere~\cite{Hanany2000,Lee2001,Stompor2001b}; their cosmological
implications are discussed in~\cite{Stompor2001a,Balbi2000,Wu2001b}.

Hereafter, we denote vectors and scalars with bold and non-bold lower
case letters (either Roman or Greek) respectively, while matrices and
operators are denoted with either bold or caligraphic upper case
letters. Vector and matrix components are indexed in parentheses,
rather than by subscript; subscripts and superscripts are used to
distinguish between different variants of a given quantity, \eg, $p$
and $t$ denote a pixel and a time domain quantity respectively. A
tilde over a quantity denotes its Fourier transform, \ie,
$$
{\bm{\tilde g }}\l(f\r)\equiv \int dt\, \bm{g}\l(t\r) \exp\l(2\pi \iota f t\r).
$$
Occasional failures of these good intentions are also acknowledged. A
summary of the most frequently used symbols is given in Table~\ref{notation:table}.

\section{Time ordered data}

\label{timedomain_main:para}

\subsection{The \maximai\ data set}

\label{maxdata:para}
\begin{figure}[t]
\vskip 0truecm
\leavevmode\epsfxsize=11.cm\epsfbox{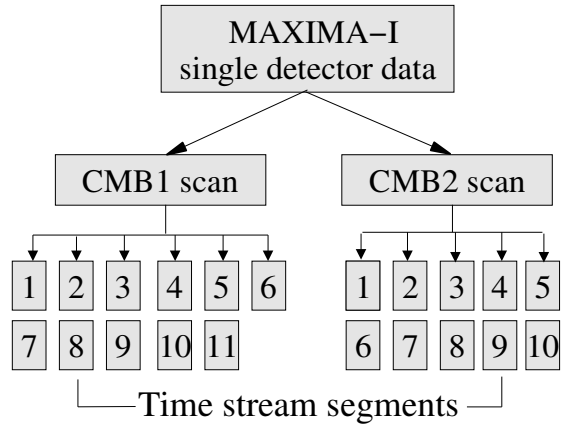}
\vskip -9truecm
\caption{A structure of the \maximai\ data set.
}
\label{maxdata:figure}
\end{figure}

The \maximai\ data and instrument have been described 
in \cite{Lee1999,Hanany2000,Lee2001}.
The \maximai\ data set consists of approximately $2,300,000$
measurements for each of 16 photometers and 4 dark channels used to
monitor the experiment. To date only data from six of the detectors
have been analyzed. These include four photometers sensitive to CMB
photons (3 with frequency bandwidths centered on $150$ GHz and 1 on
$240$ GHz), one photometer monitoring atmosphere and foregrounds at
$410$ GHz, and one `dark' bolometer (screened from incident photons)
used to search for systematic problems. Each of the data
streams for each of these detectors is divided into two parts,
hereafter called the CMB1 and CMB2 scans respectively (see Fig.~\ref{maxdata:figure}). 
These scans
were taken at different elevations and were separated in time by $\sim
1.5$ hours. Projected on the sky, they largely overlap one another,
creating a well-crosslinked map. Each of the two scans is further
subdivided into 11 (CMB1 scan) and 10 (CMB2 scan) data subsets whose
lengths range from $30,000$ to $250,000$ time samples.
 These disjoint
stretches of contiguous data -- hereafter referred to as time stream
segments -- are defined by the requirement that the noise within a segment
be approximately stationary. The noise correlations between the segments
are guaranteed to be negligible by discarding a sufficiently long
section of the time stream data between neighboring segments. In
addition, each of the segments has an overall offset and gradient
subtracted independently.

Not all measurements within a segment are to be included in the further analysis.
Measurements compromised by glitches -- cosmic rays hits, telemetry drop-outs, or other short transients -- are flagged as 'bad' and constitute gaps in a segment.
We usually determine about $2-3\%$ of the time samples as gaps.

The signal and noise are subjected to several filters
before being recorded, including a low-pass filter due to the
detector time constant and subsequent AC low- and high-pass
filters in the readout electronics. These
filters are phase-shifting (\ie, their Fourier space
representation is by complex numbers) and they define the
temporal frequency response of the instrument to a band between 
$f\simgt 0.01$Hz and $f\simlt 20$Hz. 
This frequency response together with the scan strategy 
give \maximai\ sensitivity to the 
sky features in the range of angular scales from $\sim 5$ degrees
down to $\sim 5$ arcminutes.
The instrumental
filters must be deconvolved from the time ordered data in the course
of the data analysis. The functional form of the electronic filters
is accurately measured in the laboratory. The detector time
constant is measured using  the in-flight response of the detectors
to a known source -- in the case of \maximai, Jupiter. For our
purposes here it is assumed that the instrumental filters are known to
a negligible uncertainty in the range of frequencies of interest, \ie,
$\simlt 50$Hz. (Note that the bolometer time constant has an uncertainty of
$\sim 0.5-1.0$ ms. This error is included in
the analysis of the \maximai\ data~\cite{Hanany2000,Lee2001,Stompor2001b}, but we
neglect it for the purpose of this paper.)
The data sampling interval is fixed throughout the entire observation at
$\Delta = 0.0048$ s. Hereafter, we identify an observation in the
time domain (including gaps) by a global sample-number index.

The \maxima\ scanning strategy includes an azimuthal primary mirror chop with a
frequency $\sim 0.45$ Hz superimposed on a slow azimuthal balloon gondola
oscillation with a characteristic frequency of $\sim 0.01$ Hz. Both
the position of the primary mirror with respect to the gondola and the
azimuthal position of the gondola were recorded during flight,
enabling the tracking of primary mirror and/or gondola synchronous
systematic effects.

The \maxima\ pointing is reconstructed using the observations of guide
stars made throughout the flight. With an rms accuracy better than
$\sim 1'$, pointing uncertainty is neglected during the map-making
stage, although it can, and indeed should, be included as a systematic
uncertainty of the final power spectrum (\eg,~\cite{Lee2001}).

The time stream data are assigned to pixels prior to map-making. The relatively
small angular size of the sky patch observed by \maximai\ allows us to
use simple square pixels on a grid whose rows are at constant
declination. Due to computational constraints, most of the tests
discussed in this paper have been performed using 8 arcminute pixels,
although the highest resolution \maximai\ results have
been computed using 3 arcminute pixels ~\cite{Lee2001}. 

The independence of a pixel's size on its sky position enables us to
account for the extra smoothing due to the pixelization by using a
simple, albeit approximate, pixel window function~\cite{Wu2001a}. This
approximation usually breaks down at the smallest angular scales, due
to the lack of uniformity of the sky sampling on these scales. 
Ways of alleviating this problem are discussed elsewhere~{\cite{Lee2001}.
The methods discussed below are independent of the assumed pixelization
scheme.

\subsection{Time stream modeling and simulations}

\label{timestream:para}

Drawing from the \maximai\ experience, we now explicitly list all the
features of the time stream data which we have found to be important
in the data analysis.  We also briefly explain how we incorporate
these features into our simulations, which are then used for tests of
our data analysis tools.

\subsubsection{Time stream model}

We denote the entire raw time stream from a single detector, including
the effect of the instrumental filters, as $\bm{d_F}$. As noted above,
this is subdivided into a number ($n_{seg}$) of disjoint segments, so that we can
write $\bm{d_F}\equiv \bigcup_{I=1}^{n_{seg}} \bm{d^I_F}$.

Every measurement contains contributions from both the sky and the
instrument, and is modeled as,
\begin{equation}
\bm{d_F^I}\l(i\r)=\sum_{j} \bm{F}\l(i,j\r)
\l[\bm{t}_{sky}\l(\bm{\gamma}\l(j\r)\r)+\bm{x}\l(\bm{\alpha}\l(j\r)\r)\r]+\bm{n_t}\l(i\r).
\label{simtstream}
\end{equation}
$\bm{F}$ denotes the effect of all of the instrumental filters, and for
\maxima\ is therefore a convolution of three filters -- $\bm{F}\equiv
\bm{F_{high}}\star \bm{F_{low}} \star \bm{F_{bolo}}$ -- corresponding
to the AC high-pass, AC low-pass, and bolometer low-pass filters.
$\bm{t}_{sky}\l(\bm{\gamma}\l(i\r)\r)$ is the temperature of the sky
in the direction $\bm{\gamma}\l(i\r)$ observed at time $i$.
$\bm{x}$ is any extra systematic `$\alpha$-synchronous' effect (\eg,
primary mirror chop synchronous noise) which only depends on a known
parameter $\bm{\alpha}$ (\eg, the mirror position). The dependence of
$\bm{x\l(\alpha\r)}$ on time is, therefore, exclusively a result of the
time-dependence of the parameter $\bm{\alpha}$. $\bm{n_t}\l(i\r)$
denotes the total (Gaussian, correlated) instrument noise in
observation $i$.  In fact, independent noise components are introduced
into the time stream at four different stages in the instrument and,
more precisely, the total noise is represented as
\begin{equation}
\bm{n_t}=\bm{F_{high}}\l[\bm{F_{low}} \l[\bm{F_{bolo}}\bm{n_{t1}}+\bm{n_{t2}}\r]+\bm{n_{t3}}\r]+\bm{n_{t4}}.
\label{noisedetails:eqn}
\end{equation}
Here $\bm{n_{t1}},$ $\bm{n_{t2}},$ $\bm{n_{t3}}$ and $\bm{n_{t4}}$
denote the independent noise
components added to the signal prior to the bolometer low-pass, AC high-pass,
AC low-pass filtering, and signal digitization, respectively.
 $\bm{n_{t1}}$ and $\bm{n_{t2}}$ 
components together
are expected to dominate the total instrumental noise with only
$\bm{n_{t1}}$ component displaying a $1/f$ behavior at low
frequencies attributable to temperature fluctuations of the detector.
Though all such features are of importance for proper forecasting and simulations
of a performance of an experiment,
none of these needs to be assumed in the noise estimation described
in Sect.~\ref{noisestim:para}, which directly
estimates the total noise, $\bm{n_t}$. The instrumental noise is
assumed to be Gaussian and stationary within each segment, and is
described by a (segment-dependent) noise power spectrum, $P\l(f\r)$. Each segment is
assumed to consist of measurements evenly spaced in time.  However,
the data constituting gaps are not to be included in a final map.

\subsubsection{Simulations}

In order to test our data analysis pipeline we want to be able to
simulate the time stream of a \maximai\ like experiment. The
simulation is designed to incorporate all the important features of
the actual data set listed above, including the gap and segment
structure, scanning strategy, and an approximate (symmetric)
beam~\cite{Wu2001a}.

The simulated time stream is described by \eqn~(\ref{simtstream}).
The sky signal ($\bm{t_{sky}}$) is generated by applying the known
pointing solution to a simulated CMB sky, generated as a
high-resolution Gaussian realization given some fiducial cosmological
parameters. 
We also include a primary mirror chop synchronous systematic effect,
$\bm{x\l(\alpha\r)}$, in our simulations, varying both its functional
dependence on the primary mirror position and its amplitude.
The instrumental filters are then applied to the simulated
time stream as described by \eqn~(\ref{simtstream}). Their functional
form follows that of actual \maxima\ filters.

Finally we add Gaussian correlated noise to each time sample. This is
modeled according to \eqn~(\ref{noisedetails:eqn}) assuming that each
component, $\bm{n_{ti}}$, has a power spectrum in a form given by,
\begin{equation}
P_{sims}\l(f\r)=\sigma_{sims}^2\l(1+\frac{f_{knee}}{f}\r).
\label{nspec}
\end{equation}
This approach means that our simulations mimic the range of floating
point operations required in analyzing the real data, giving us some
insight into possible numerical error accumulation in our
data analysis pipeline.

\subsection{Time stream processing and noise estimation}

\label{noisestim:para}

\begin{figure}[t]
\vskip 0.0truecm
\leavevmode\epsfxsize=8.5cm\epsfbox{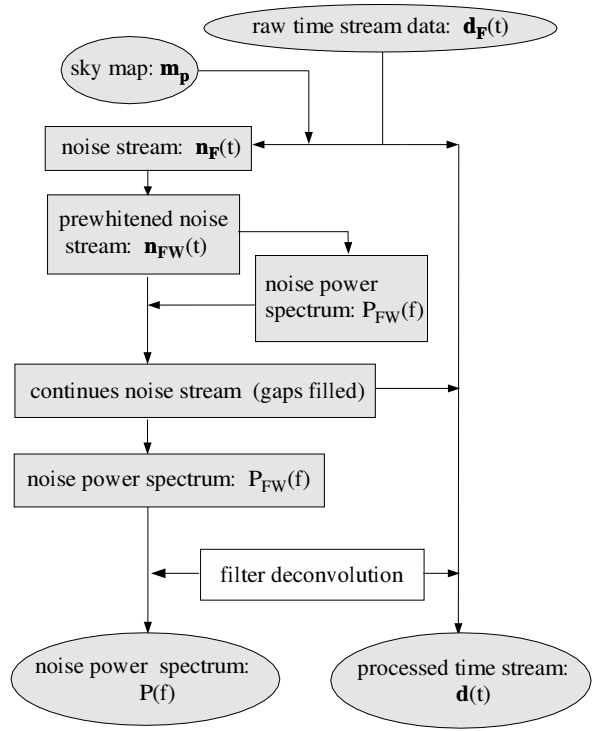}
\vskip -1.5truecm
\caption{A flow-chart of the noise estimation procedure outlined in
Sect.~\ref{noisestim:para}. Objects inscribed in ovals are either
input or output products of the procedure. This procedure can be incorporated
as a part of the iterative noise estimation as described in Sect.~\ref{iteration:para}.
In this case, its products (noise power spectrum and a processed time stream) 
are used to generate a sky map, which is one of input objects as shown in this chart.
}
\label{noisechart:figure}
\end{figure}

We now describe the procedure for simultaneously estimating the
time-domain noise power spectrum and restoring the stationarity of
time stream noise by re-filling the flagged-data gaps in the time
stream.

Typically, map-making methodologies assume that the time-domain noise
power spectrum is precisely known, even though in practice it has to
be estimated from the time stream data itself. The error in this
estimation is therefore not included in general in the end-to-end data
analysis error budget (although see \cite{Ferreira2000} for a possible
way of tackling this issue). This fact clearly highlights the need for
a high level of precision at this stage.

\begin{figure*}[t]
\vskip -4truecm
\leavevmode\epsfxsize=18.cm\epsfbox{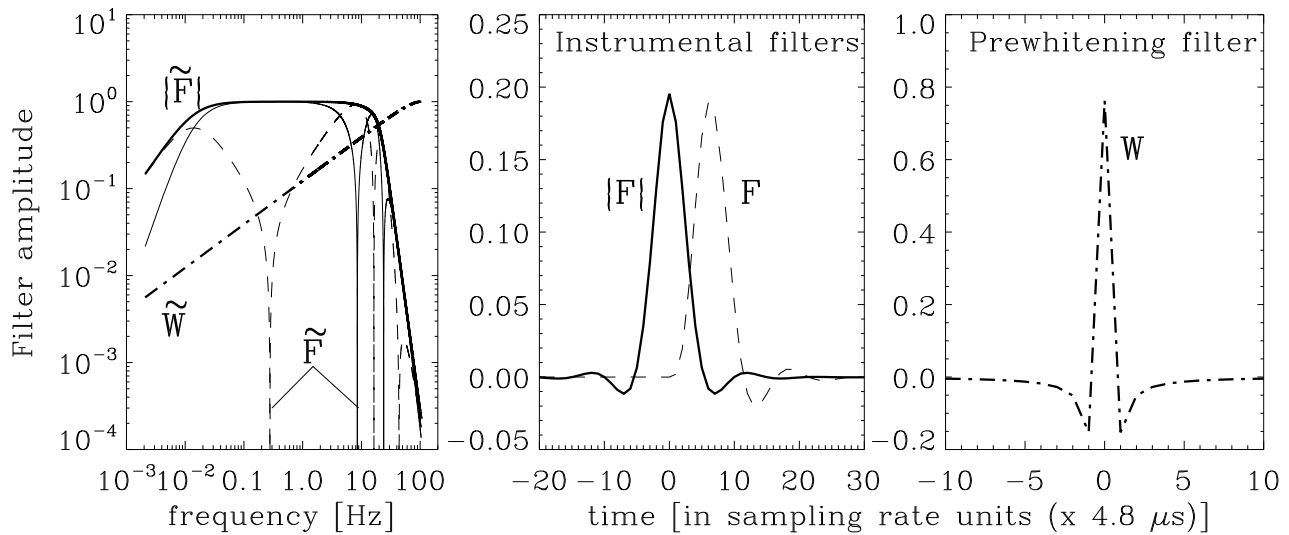}
\vskip -2truecm
\caption{Filters used in the analysis of \maximai\ data. Left
panel shows a frequency domain representation of the filters.  Thin
solid and dashed lines show the (absolute values of) real and imaginary parts of the total
instrumental filter, $\wtilde F\l(f\r)$, and the thick line its complex
amplitude, $|\wtilde F\l(f\r)|$.  The dot-dashed line depicts the real
prewhitening filter, $\wtilde W\l(f\r)$, as defined in
\eqn~(\ref{whfilter:eqn}) ($2\beta=1.0$). The time domain representation of these
filters shown in middle panel: solid line - total instrumental 
filter and dashed line its amplitude; and right panel - the
prewhitening filter. The effective time domain widths of these filters
assumed in the analysis are 100 time samples for both instrumental
filters and 50 time samples for the prewhitening filter.}
\label{filters:figure}
\end{figure*}

The time stream model described by \eqn~(\ref{simtstream}) is quite
complex. Some of its features are of less importance for this Section
so, for simplicity, we start by assuming that the time stream data
contains only Gaussian noise (\ie, no sky or systematic signals,
$\bm{t_{sky}}, \bm{x} \ll \bm{n_t}$) and focus on resolving problems
in the noise estimation arising from the presence of gaps in the time
stream and of noise correlations on time scales comparable to the
length of an entire data segment.

The presence of discontinuities (gaps) in the time stream poses a
two-fold problem. It is an obstacle both to estimating the time stream
noise spectrum and to devising a fast and efficient map making
algorithm (see Sect.~\ref{mapmaking_main:para}). It is therefore desirable to
restore time stream continuity without compromising the stationarity
of the noise or sacrificing too much of the data. These two goals,
estimating the noise power spectrum and restoring the continuity of the
time stream, are achieved by a procedure described below 
(see Fig.~\ref{noisechart:figure} for its synopsis).

\subsubsection{A pure noise time stream}

The initial input at this stage consists of the time stream data with
the instrumental filters convolved, $\bm{d_F}$.  The filters can be
deconvolved in the frequency domain,
\begin{equation}
\bm{d}\l(t\r)=\int df\,{\bm{\tilde d_F}}\l(f\r) \wtilde{F}^{-1}\l(f\r) \exp\l(-2\pi\iota f t\r).
\end{equation}
This deconvolution could be done immediately, prior to any further
time stream manipulation, but this is not recommended due to the 
high-frequency noise it adds to the time stream (see
Figs.~\ref{filters:figure} \&\ \ref{noisestim:figure}). Although this noise would not be a
problem for our formalism in theory, it can be a source of significant
numerical error in any practical implementation, possibly biasing the
final results. This numerical error is usually incurred while Fourier
transforming the noise power spectrum. Instrumental filter deconvolution
causes the noise power at the highest frequencies to dominate the
total power of the time stream noise by many orders of magnitude, and
consequently a small error in the total noise power estimate
translates into a substantial error in the noise power estimate in the
lower range of frequencies of most interest.  The remedy is either to
leave the instrumental filter deconvolution until after the noise
estimation stage, or to perform the deconvolution at the beginning but concurrently 
convolving an extra (real) filter, $\wtilde{X}_{re}(f)$, to suppress the
high frequency noise, i.e.,
\begin{equation}
\bm{d}\l(t\r)=\int df\,{\bm{\tilde d_F}}\l(f\r) \wtilde{X}_{re}\l(f\r) \wtilde{F}^{-1}\l(f\r) \exp\l(-2\pi\iota f t\r).
\end{equation}
In this case, a conventional choice is to
apply a real filter given in the frequency domain as the amplitude of
the total instrumental filter, \ie, $\wtilde{X}_{re}\l(f\r)=\l|\wtilde F\l(f\r)\r|$.  In some
map-making algorithms this filter can be deconvolved from the time
stream data once the noise power has been estimated and the noise
stationarity restored so that effectively no filtering has been
applied to the time stream data due the course of data analysis. This
is the case when the noise correlation matrix in the time domain does not
have to be computed explicitly, as in, \eg, the
approximate map-making algorithms discussed below.  In these cases the
precise shape of the extra filter is not important. If, however, a
full noise correlation matrix in time domain is needed, as in
exact map-making algorithms, such a deconvolution should be
avoided. In this case the precise form of the extra filter is very
important, and should be designed to suppress the high frequency noise
power sufficiently, while leaving the widest possible range of lower
frequencies unaffected.

\begin{figure*}[t]
\vskip -4truecm
\leavevmode\epsfxsize=18.cm\epsfbox{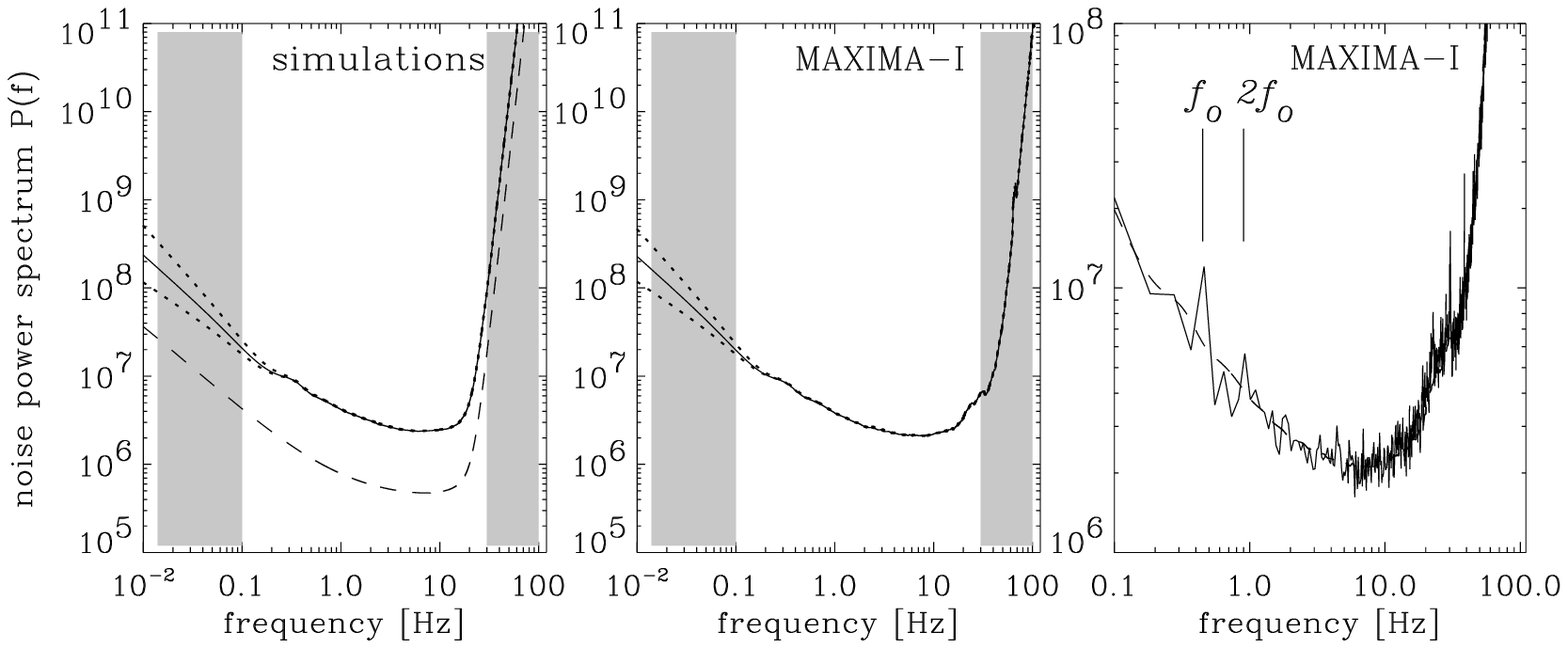}
\vskip -2truecm
\caption{Estimated noise power spectra for simulated and real data.
In left panel, the dashed line shows an effective noise power spectrum with $2\beta=1.0$
used for a simulation (its amplitude is scaled down by a factor of 5
to avoid overlapping with solid line).  Solid and dotted lines show
noise power spectra recovered from this simulation using a
prewhitening filter as given by \eqn~(\ref{whfilter:eqn}) with
$2\beta=1.0$ - solid line; $2\beta=1.25$ (top) and $2\beta=0.75$
(bottom) dotted lines.  Middle panel shows noise power spectra for a
single segment of the actual \maximai\ data. Again the recovery has been
made using various prewhitening filters as in the left panel.  In all
the results displayed in these panels aggressive smoothing have been
applied to facilitate better comparison. For actual calculations, such
an approach does not have to be the best option.  Right panel shows a
noise spectrum obtained just as a result of averaging at the second
step of the noise estimation procedure, prior to any smoothing. Two
small spikes at $\sim 0.45$Hz and $0.9$Hz correspond to a primary
mirror chop frequency and its first harmonic. The overplotted dash
line is a smoothed version of this spectrum as shown in the middle
panel with a solid line.  Shaded areas in both left and middle panels show a
region of frequencies we marginalize over to avoid either spurious
high frequency noise or a noise estimation uncertainty at the low
frequency end.  }
\label{noisestim:figure}
\end{figure*}

In the following we set the initial width of all the gaps to be no
less than twice the effective width of the instrumental filters (which
we conservatively choose to be 100 time samples, see
Fig.~\ref{filters:figure}), reflecting our ignorance about the
origin of some of the gaps. Similarly, we always widen gaps on both
sides by an effective filter width when (de)convolving any filter from
the raw time stream.

The noise estimation algorithm described here is applied to each time
stream segment separately and consists of the following steps:

\begin{enumerate}

\item{{\sl Reducing the noise correlation length, $\lambda_c$:}\\ 
The time stream segment is convolved with a prewhitening filter,
\begin{equation}
\bm{\tilde{d}_{FW}}\l(f\r)\equiv \wtilde{W}\l(f\r)\bm{\tilde{d_F}}\l(f\r),
\end{equation}
and simultaneously the gaps are widened to account for the width of
this filter. The prewhitening filter ($\wtilde W$) is usually a
half-differencing real filter~\cite{Tegmark1997design}:
\begin{equation}
\wtilde W\l(f\r)\equiv \sin^\beta\l(\phi/2\r),\label{whfilter:eqn}
\end{equation}
with a phase $\phi = \pi f \Delta$ and a sampling interval $\Delta$. The
prewhitening index, $\beta$, reflects the expected low frequency
behavior of a noise power spectrum, \ie, $P\l(f\r)\sim f^{-2\beta}$
for $f\rightarrow 0$. It is adjusted for each segment separately
(Fig.~\ref{filters:figure}), so that the prewhitened power spectrum, $P_W\l(f\r)\sim \mbox{{\sc const}}$
for $f\rightarrow 0$.
The functional form of the prewhitening filter is tested in the
subsequent stages of the data analysis, and, if necessary, it may be
refined and the entire procedure repeated.}

\item{{\sl Estimating the noise power spectrum of the prewhitened
time stream, $P_{FW}\l(f\r)$:}\\ 
Sections of the time stream segment are located that have no gaps, are
longer than the expected noise correlation length, and are separated
from each other by at least this distance. These are used to compute a
series of statistically independent estimates of the noise power
spectra using standard FFT-based methods (\eg, \cite{Press1992}).
These estimates are then checked for stationarity and, if consistent
with one another, averaged. If they do not exhibit stationarity then
the original time stream segments have to be redefined and made
appropriately shorter. The final -- average -- power spectrum of the
stationary noise in a segment can be additionally smoothed and
extra(inter)polated to the frequencies corresponding to the full
segment.}

\item{{\sl Restoring time stream continuity:}\\ 
The gaps are filled with a constrained realization of the
noise~\cite{Hoffman1991} which is assumed to be Gaussian with 
a power spectrum, $P_{FW}\l(f\r)$, as determined on the previous step
 and corresponding
noise correlation matrix (see \eqn~\ref{nt}), denoted as $\bm{N_t}$
below.  The noise within each gap is constrained only by good data
samples in its vicinity, \ie, those which are within the noise
correlation length, $\lambda_c$, of the gap edges,
\begin{eqnarray}
&&\bm{d_{FW}}\l(i\r) = \bm{\xi}\l(i\r)+\\
&&+\sum_{m,n}\bm{N_t}\l(i,m\r)\bm{N'_t}^{-1}\l(m,n\r)\l(\bm{d_{FW}}\l(m\r)-\bm{\xi}\l(m\r)\r).\nonumber
\end{eqnarray}
Here, $\bm{N'_t}$ is a $2\lambda_c \times 2\lambda_c$ matrix
describing noise correlations between the time samples just outside of
a given gap, $\bm{\xi}$ is a vector of Gaussian, correlated (but
unconstrained) random variables, and $i$ indexes the time samples
within a gap.  The sum is over $m$ and $n$, both of which index the
time samples outside of a given gap which are being used to constrain
the noise realization within the gap. Due to required inversion of the
matrix $\bm{N'_t}$ the computational cost of the procedure is ${\cal
O} \l(\lambda_c^3\r)$ per gap, and it is therefore crucial that
$\lambda_c$ is sufficiently short.}

\item{{\sl Re-estimating the noise power spectrum:}\\ 
The entire time stream segment, with gaps filled, is now used to
re-estimate the noise power spectrum. At this stage the low frequency
behavior of the noise can be tested; if it is found to be
significantly non-flat (\ie, non-white), or to differ from the
spectrum estimated at stage 2, then the prewhitening filters used at
the stage 1 must be adjusted and the entire sequence repeated.  }

\item{{\sl Deconvolving the filters:}\\
Once the noise power spectrum has been estimated, and the gaps filled
with the constrained noise realization, the instrumental and
prewhitening filters can be deconvolved from both the time stream and
the noise power spectrum. In general, this applies to both the
instrumental (or $|\wtilde{X}_{re}\l(f\r)|$) and prewhitening filters. As
mentioned above, the deconvolution of the instrumental filters is not
always advisable since it may introduce numerical errors. The
deconvolution of the prewhitening filter is also sometimes postponed,
and accounted for only in the algebra of the map-making algorithm
~\cite{Tegmark1997design}. Such an approach attempts to make use of
the shortened correlation length of the prewhitened noise to cut the
number of floating point operations required to make the map. In
practice we find that it is difficult to take advantage of this in any
realistic map-making algorithm (see Sect.~\ref{mapmaking_main:para}),
while, as we argue below, performing the deconvolution at the outset
of the noise estimation procedure helps to alleviate a number of
problems.}

\end{enumerate}

This procedure attempts to estimate the ensemble average noise power
spectrum using just one realization of the time stream segment. This is
clearly not enough, especially in the presence of correlations on
time scales comparable with the length of the segment. Prewhitening
helps to alleviate this problem, yet it requires some extra knowledge
about the functional form of the prewhitening filter. In this
approach, an educated guess, followed by iterative refinement, is then
tested {\sl a posteriori}. For a prewhitening filter of the form
given in \eqn~(\ref{whfilter:eqn}), we usually find that the power-law
index $\beta$ can be uniquely determined with an error not bigger than
$|\Delta\beta|\simlt 0.1$ by comparing the power spectra computed at
the 2nd and 4th stages of the procedure (see
Fig.~\ref{noisestim:figure}).

Computing and averaging the noise power spectra of the independent,
gap-free, sections of a segment (step 2 above) helps to recover a better ensemble
average spectrum at intermediate and high frequencies. Smoothing in
the frequency domain can
also be applied to get even closer to the ensemble average.
However, in both cases we need to assume that the real noise power spectrum
does not display any significant variation on scales smaller than the
smoothing or frequency re-sampling scale.

The result of this procedure is an estimate of the ensemble average
noise power spectrum in the time domain which is reliable at
sufficiently high frequencies (for \maximai\ for $f\simgt 0.1$Hz), but
sample-error dominated at the low frequency end ($f\simlt 0.1$Hz).
To minimize the effect of the low frequency uncertainty, in the
subsequent analysis we marginalize over the low frequency part of the
spectrum (see Sect.~\ref{marginalization:para}).

The entire procedure restores the continuity of the time stream
segments: the noise is stationary over an entire segment, including gaps,
and the sky signal in gaps is zero.  
This is important in the subsequent stages of the data analysis; we
might worry that adding a random (albeit constrained) signal to the
data introduces an extra degree freedom and undermines the uniqness of
the result. Although this is true, while different realizations of the
random component may indeed lead to slightly different results, all of
these results are necessarily statistically equivalent, with no bias
introduced. Moreover, the impact of the extra randomness is
significantly reduced by the gap pixel marginalization described in
Sects.~\ref{extrapixels:para} \&\ ~\ref{marginalization:para}. Consequently, we find that, for
\maximai\, the final maps and their power spectra are robust and not
affected by the random element of this procedure.

Generating constrained noise realizations to fill the gaps is the most
time consuming part of this procedure. How important this is for the
consistency and accuracy of the entire data analysis pipeline depends
on the frequency, size, and regularity of the gaps in a time stream
segment. In the case of a handful of narrow, randomly scattered, gaps,
an unconstrained, uncorrelated, random noise realization with an rms
determined from the rest of the time stream may serve as a convenient
first approximation for the analysis.

\subsubsection{A time stream with non-negligible sky signal}

If the sky signal present in the time stream data is not negligible,
but still sub-dominant,
then the noise estimation has to be performed iteratively, following
an algorithm proposed by Ferreira \&\ Jaffe~\cite{Ferreira2000}. 
In this case we distinguish between
the full (noise + signal) time stream and the noise-only time stream,
where the latter is the noise in the time stream data. 
At each step of the iteration, 
a maximum-likelihood map estimated on the previous step is subtracted from the time stream
giving the current best estimate of the noise stream (see Sect.~\ref{iteration:para}).
The above noise power spectrum procedure is
applied in full only to the noise time stream; only the instrumental
filters deconvolution and related gap-widening are performed on the
actual time stream (see Fig.~\ref{noisechart:figure}).
Once the noise estimation has been accomplished, only the noise stream continuity is restored.
Therefore the gap
time samples from the noise time stream data are used to replace their
analogues in the full (unprocessed) time stream. This avoids unnecessarily wasting
or biasing good data samples which happen to be in a vicinity of a
gap.

A simple extension of this iterative approach also allows us to
account for problems related to presence of synchronous systematic
signals in the data. We discuss the appropriate algorithms and related
issues in some detail in Sect.~\ref{iteration:para}.

\subsection{The time domain noise correlation matrix}

\label{timecorr:para}

Formally, maximum likelihood map-making requires the full time-time
noise correlation matrix, rather than just a noise power spectrum.
For an idealized time stream the former is just the Fourier transform
of the latter,
\begin{equation}
\bm{N_{Ct}}\l(i,j\r)\equiv \int df\, P\l(f\r)\exp\l(-2\pi\iota f (i-j)\Delta\r).
\label{nct}
\end{equation}
For a finite time stream this expression leads to a circulant matrix
denoted by the subscript $\bm{C}$~\cite{Golub1983}. If the correlation length in the
time stream is less than half of the time stream length, then an
alternative approximate correlation matrix, $\bm{N_t}$, is given by
\begin{equation}
\bm{N_t}\l(i,j\r)\simeq \l\{
\begin{array}{l l}
\dsp{\bm{N_{Ct}}\l(i,j\r);} & \dsp{\mbox{if } |i-j| < \lambda_c,}\\
\dsp{0;}                    & \dsp{\mbox{otherwise.}}
\end{array}
\r.
\label{nt}
\end{equation}
In practise, numerical error in the estimation (and Fourier transform)
of the noise power spectrum means that $\bm{N_t}$ computed in this way
may not even be positive definite. Such problems can be alleviated by
the introduction of an extra power spectrum smoothing window,
$S\l(f\r)$, designed to truncate smoothly the correlations once their
amplitude reaches the limits of numerical accuracy. Again, it is
prudent to apply the window function prior to deconvolving the
prewhitening filter. \eqn~(\ref{nct}) then assumes a somewhat
more complicated form,
\begin{eqnarray}
\bm{N_{Ct}}\l(i,j\r) & = & \int df\, |\wtilde W\l(f\r)|^{-2} \exp\l(-2\pi \iota f (i-j)\Delta\r)\nonumber\\
                & \times & \int df'\, S\l(|f-f'|\r) P_W\l(f'\r), \label{nctfull}
\end{eqnarray}
where $\wtilde W\l(f\r)$ is given by \eqn~(\ref{whfilter:eqn}). In
the following, we make use of this expression whenever computing
$\bm{N_{Ct}}$. Its inverse is approximated by an analogous
formula~\cite{Golub1983},
\begin{eqnarray}
\bm{N_{Ct}}^{-1}\l(i,j\r)&=& \int df\, |\wtilde W\l(f\r)|^{2} \exp\l(-2\pi \iota f (i-j)\Delta\r)\nonumber\\
&\times&\int df'\, S\l(|f-f'|\r) P_W^{-1}\l(f'\r). \label{invnctfull}
\end{eqnarray}
Our usual choice for $S(f)$ is a Gaussian with an appropriately tuned
width (typically of $1000-5000$ time samples for \maximai).

\section{Map-making: formalism}

\label{mapmaking_main:para}

\subsection{The basic framework}

\label{mapmaking:para}

The procedure described above provides the input for the map-making
{\em per se}; here, we take as given the time stream data with
instrumental filters deconvolved and gaps filled, and the corresponding
noise power spectrum for each segment.

In this Section we consider a case of a somewhat idealized time
stream, leaving a discussion of the fully realistic case to Sect.~\ref{amendments}.  
We consider a time stream consisting of a single
segment, neglect most of the systematic contributions, but do admit the
presence of (now filled) gaps in the data. The simplified \eqn~(\ref{simtstream}) reads now,
\begin{equation}
\bm{d_t}=\bm{A}\bm{m_p}+\bm{n_t}. \label{simpletimestream}
\end{equation}
Here $\bm{A}$ is a pointing matrix assigning each time sample to an
appropriate pixel (or a set of pixels in a case of differencing experiments)
on the sky (as observed at the given time) with the
sky signal given by a pixelized map: $\bm{m_p}$. $\bm{n_t}$ is the
(Gaussian) noise time stream with correlations given by the power
spectrum estimated as in the previous Section.

In this case we can write a closed form solution for the
map~\cite{Wright1996a,Tegmark1997design} (but see~\footnote{We have
omitted the extra (prewhitening) filter matrix, $\bm{D}$, introduced in~\cite{Tegmark1997design}
to reduce the number of floating point operations in the matrix
multiplications of \eqn~(\ref{mapmaking}). In fact the
band-diagonality of `prewhitened' matrices (\ie, $\bm{M'}\equiv\bm{D}\bm{M}\bm{D}^T$ and
$\bm{N_t'}^{-1}\equiv \bm{D}\bm{N_t}^{-1}\bm{D}^T$) can be exploited only if
the products in \eqn~(\ref{noisemap}) are performed from outside
inwards, and though that may speed up computation of $\bm{A}^T\bm{M'}$
the computational cost of the subsequent products,
\ie, $(\bm{A}^T\bm{M'})\bm{N_t'}^{-1}$ and
$(\bm{A}^T\bm{M'})\bm{N_t'}^{-1}(\bm{A}^T\bm{M'})^T$, offsets any
advantages on the earlier stage.}). The maximum likelihood estimates of the map,
$\bm{m_p}$, and the pixel-pixel noise correlation matrix, $\bm{N_p}$,
then read,
\begin{eqnarray}
\bm{m_p}&=&\l(\bm{A^T}\bm{M}\bm{A}\r)^{-1}\,\bm{A^T}\bm{M}\bm{d_t},\label{mapmaking}\\
\bm{N_p}&=&\l(\bm{A^T}\bm{M}\bm{A}\r)^{-1}\l(\bm{A^T}\bm{M}\bm{N_t}\bm{M}\bm{A}\r)\l(\bm{A}^T\bm{M}\bm{A}\r)^{-1}.\label{noisemap}
\end{eqnarray}
Here, $\bm{M}$ is a positive-definite symmetric matrix, and $\bm{N_t}$
the time domain noise correlation matrix (\eqn~\ref{nt}). If
$\bm{M}=\bm{N_t}^{-1}$, then \eqns~(\ref{mapmaking}) \&\ (\ref{noisemap})
give minimum variance estimates of $\bm{m_p}$ and $\bm{N_p}$. Other
choices trade a larger variance for increased computational speed. In
particular Tegmark~\cite{Tegmark1997design} proposed the choice for
$\bm{M}$ of the circulant part of the $\bm{N_t}$ matrix. We discuss
this option in Sect.~\ref{circmaps:para} below.

In the following we present some different approaches and their
application to a realistic \maxima-like data set.

\subsection{Minimum variance approaches}

\label{minvarmaps:para}

\subsubsection{Exact methods}

With $\bm{M}=\bm{N_t}^{-1}$, we get the minimum variance estimates,
\begin{eqnarray}
\bm{m_p}&=&\l(\bm{A^T}\bm{N_t}^{-1} \bm{A}\r)^{-1}\,\bm{A^T}\bm{N_t}^{-1}\bm{d_t},\label{minvarmap}\\
\bm{N_p}&=&\l(\bm{A^T}\bm{N_t}^{-1}\bm{A}\r)^{-1}.\label{minvarnoise}
\end{eqnarray}
The exact implementation of these equations seems a daunting task. The
time stream length may easily reach many tens and hundreds of million
of samples, making exact inversion of the time domain noise
correlation matrix prohibitive. However, for a \maximai\ like
experiment with segments lengths reaching only \order{10^5}
samples an exact implementation is feasible on a moderately powerful
workstation.

At the core of the implementation lies the observation that for a
gap-free time stream of the length $n_s$ with stationary noise the time domain noise
correlation matrix is Toeplitz, with
$\bm{N_{t}}\l(i,j\r)=\bm{N_{t}}\l(\l|i-j\r|,0\r)$. The inversion of a
Toeplitz matrix can be performed in as few as \order{n_s^2} operations --
clearly feasible for \order{10^5} time samples -- and even
faster algorithms are possible (\eg, \cite{Golub1983}) bringing that
number down to \order{n_s\log^2 n_s}.  The number of operations can be
further reduced if the noise correlation length is shorter than the
time stream segment length and the noise correlation matrix is therefore
band-diagonal.

If gaps are present in the time stream then, in principle, the
Toeplitz (stationary) character of the time domain noise correlation
matrix is lost.  However the gap-filling procedure described above is
explicitly designed to restore the stationarity of the noise in the
time stream. Since the gap samples contain no sky signal (by
construction) they can be treated as data taken while observing a
fictitious signal-free pixel in the sky map $\bm{m_p}$. Once the map
and its noise matrix are estimated, this extra gap pixel is rejected
from the map and the pixel domain noise correlation matrix
(effectively marginalizing over it). This is a special application of
the extra pixel method discussed in detail below (see Sect.~\ref{extrapixels:para} 
and also ~\cite{daCosta1999}).

The computational effort then scales with a number of pixels $n_{pix}$
and a number of time samples $n_s$ as:
\begin{itemize}
\item{noise inverse in time domain: \\
\hskip2.5cm $\bm{N_t} \longrightarrow \bm{N_t}^{-1}$ : \order{n_s^2};}
\item{noise inverse in pixel domain: \\ 
\hskip2.5cm $\bm{N_t}^{-1}, \bm{A} \longrightarrow \bm{A}^T \bm{N_t}^{-1} \bm{A}$ : \order{n_s^2};}
\item{noise weighted map: \\
\hskip2.5cm $\bm{N_t}^{-1}, \bm{A}, \bm{d_t} \longrightarrow \bm{A}^T \bm{N_t}^{-1} \bm{d_t}$ : \order{n_s^2};}
\item{noise matrix in pixel domain: \\
\hskip2.5cm $\bm{N_p}^{-1} \longrightarrow \bm{N_p}$ : \order{n_{pix}^3};}
\item{final map: \\
\hskip2.5cm $\bm{N_p}, \bm{A}^T \bm{N_t}^{-1} \bm{d_t} \longrightarrow \bm{N_p} \l(\bm{A}^T \bm{N_t}^{-1} \bm{d_t}\r)$ : \order{n_{pix}^2}.}
\end{itemize}

\begin{figure*}[t]
\vskip -5 truecm
\epsfxsize=19.cm\epsfbox{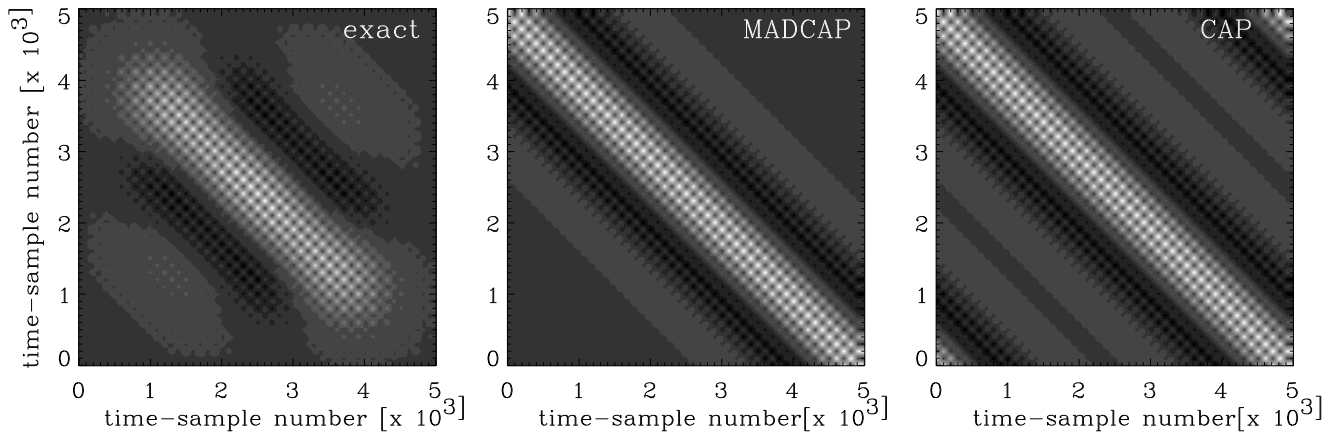}
\vskip -2.5truecm
\caption{Inverse time-domain noise correlation matrices
computed using MADCAP (middle panel) and circulant (right)
approximations and compared with the exact result (left). For
visualization purposes the time-stream
segment length assumed here is only $5,000$ samples and the
correlation length, $\lambda_c\simeq 1000$.
}
\label{tnoise:fig}
\end{figure*}

For the first three items substantial savings can
be made if the inverse time-time noise correlation matrix,
$\bm{N_t}^{-1}$, is assumed
to be band-diagonal -- an approximation usually well fulfilled for the
inverse of a band-diagonal noise correlation matrix.  If this is the
case, and, in addition, the time domain correlation length is short,
then the inverse pixel-pixel noise correlation matrix, $\bm{N_p}^{-1}$, is often rather
sparse. In this case additional savings can be made by using the
sparse matrix algorithms, \eg, Sherman-Morrison-Woodbury formula (\eg, \cite{Press1992}) to calculate both the
map and the noise correlation matrix in the pixel domain. 

The memory requirements are also reduced by considering the various
matrices' structure. Since we only need to keep the first row of the
Toeplitz $\bm{N_t}$ matrix and we do not need $\bm{N_t}^{-1}$ explicitly,
the major memory requirements are set
by the size of the $\bm{A}^T\bm{N_t}^{-1}$ matrix, which is
\order{n_{pix}n_s}, and of the noise correlation for the map, \order{n_{pix}^2}. 
Note that the first of these limits can be reduced at the
expense of the operation counts, since there is no need to save all
$n_{pix} n_s$ elements of this matrix provided we are prepared to
re-calculate them as needed.

\subsubsection{Approximate methods}

One way to speed up map-making, while at the same time providing a
good approximation to the optimal minimum variance map was proposed by
Ferreira and Jaffe~\cite{Ferreira2000} and incorporated into the
MADCAP package by Borrill~\cite{Borrill1999mad}. Rather than explicitly
inverting the $\bm{N_t}$ matrix we instead use the approximation,
\begin{equation}
\bm{N_t}^{-1}\l(i,j\r)\simeq \l\{
\begin{array}{cl} 
\bm{N_{Ct}}^{-1}\l(i,j\r), & \mbox{\rm if }|i-j| \le \min\l(n_s/2,\lambda_c\r), \\
0, & \mbox{\rm else.} \\
\end{array}
\r.
\label{FJapprox}
\end{equation}
Here $n_s$ and $\lambda_c$ are the time stream and correlation lengths
respectively, and the subscript $\bm{C}$ denotes the circulant part of
the noise correlation matrix (\eqn~\ref{nctfull}). Apart from this
approximation, the remaining steps follow precisely as before.
Hereafter we refer to this approach as the MADCAP approximation.

Inversion of a circulant matrix can be accomplished using FFTs
(\eqn~\ref{invnctfull}) in only \order{n_s\log n_s} operations.  This
approach is therefore designed to reap the benefits of Fast Fourier
methods, while at the same time providing a good approximation to the
exact solution.  If $n_s \gg 4 \lambda_c$ the fraction of
$\bm{N_t}^{-1}$ matrix elements seriously misestimated (\ie, affected
by the segment `edge' effects, see Fig.~\ref{tnoise:fig}) by this
procedure should only be \order{4\lambda_c^2/n_s^2}, and
its performance in the large $n_s$ limit should be
satisfactory. However, because the inverse of a Toeplitz matrix is
Teoplitz only when it is also circulant this is the only case when
this approach is exact. We demonstrate how well this approximation
fares in reproducing the pixel domain map and noise correlation matrix
in realistic cases below (Sect.~\ref{assesment:para}).

The operation count changes only for the two items in the list, which
now read,
\begin{itemize}
\item{noise inverse in time domain: \\ 
\hskip2.5cm $\bm{N_t} \longrightarrow \bm{N_t}^{-1}$ : \order{n_s \log n_s};}
\item{noise weighted map: \\
\hskip2.5cm $\bm{N_t}^{-1}, \bm{A}, \bm{d_t} \longrightarrow \bm{A}^T
\bm{N_t}^{-1} \bm{d_t}$ : \order{n_s \log n_s}}.
\end{itemize}
The scaling of the noise weighted map operation count above assumes the use
of FFTs for the Toeplitz matrix multiplication -- a trick explained
in~\cite{Golub1983}. 
The memory requirement is set either by the larger of the size of the
noise matrix in the pixel domain $n_{pix}^2$ or the time
stream length $n_s$. 
If an efficient, fast --
${\cal O} \l(n_s\log^2 n_s\r)$ -- implementation of the Toeplitz matrix
inversion is viable then, at least in the common case of an
approximately band-diagonal inverse noise correlation matrix $\bm{N_t}^{-1}$,
the major computational advantage of the
approximate method over the exact one would vanish, and the memory
requirement would remain as its only asset.

\subsection{Circulant approaches}

\label{circmaps:para}

If we take $\bm{M}=\bm{N_{Ct}}^{-1}$ then the noise correlation matrix
in the pixel domain can be expressed as~\cite{Tegmark1997design},
\begin{eqnarray}
\bm{N_p}&=&\bm{N_{Cp}}+\bm{N_{Sp}},\nonumber\\
\bm{N_{Cp}}&\equiv&\l(\bm{A}^T\iNct \bm{A}\r)^{-1},\label{circnoise1}\\
\bm{N_{Sp}}&\equiv&\bm{N_{Cp}}\l(\bm{A}^T\iNct \bm{N_{S t}}\iNct \bm{A}\r)\bm{N_{Cp}},\nonumber
\end{eqnarray}
where the time domain noise correlation matrix has been decomposed
into its circulant ($\bm{N_{C t}}$) and sparse ($\bm{N_{S t}}$) parts,
\begin{equation}
\bm{N_t}\equiv \bm{N_{C t}}+\bm{N_{S t}}.\label{circnoise2}
\end{equation}
The sparse term compensates for the off-diagonal corners of the
circulant matrix. As in the case of
the MADCAP approximation, the only inversion required in the time
domain is that of a circulant matrix. Hence it has been argued
that, because the $\bm{N_{S t}}$ matrix
should be very sparse, the operation count in this approach should be
significantly lower than in the exact minimum variance approach~\cite{Tegmark1997design}.

The disadvantage of this approach is not in the non-minimum-variance map that
results (since the loss of precision is usually insignificant) but
rather in implementing its more complicated algebra. If no assumption
is made about band-diagonality, then the computational costs scale as:
\begin{itemize}
\item{noise inverse in time domain: \\
$\bm{N_{Ct}} \longrightarrow \bm{N_{Ct}}^{-1}$ : \order{n_s\log n_s};}
\item{noise inverse in pixel domain: \\
circulant part: \\
$\bm{N_{Ct}}^{-1}, \bm{A} \longrightarrow \bm{A}^T \bm{N_{Ct}}^{-1} \bm{A}$ : \order{n_s^2}; \\
sparse part: \\
$\bm{N_{S t}},\bm{N_{Ct}}^{-1} , \bm{A} \longrightarrow$ \\
\phantom{} \hskip1.0cm $\bm{A}^T\bm{N_{Ct}}^{-1}\bm{N_{S t}}\bm{N_{Ct}}^{-1}\bm{A}$ : \order{n_{pix} n_s^2};
}
\item{noise weighted map: \\
$\bm{N_{Ct}}^{-1}, \bm{A}, \bm{d_t} \longrightarrow \bm{A}^T \bm{N_{Ct}}^{-1} \bm{d_t}$ : \order{n_s \log n_s};}
\item{noise matrix in pixel domain: \\
$\bm{N_p}^{-1} \longrightarrow \bm{N_p}$ : \order{n_{pix}^3}.}
\end{itemize}
\begin{figure*}
\vskip -4truecm
\leavevmode\epsfxsize=18.cm\epsfbox{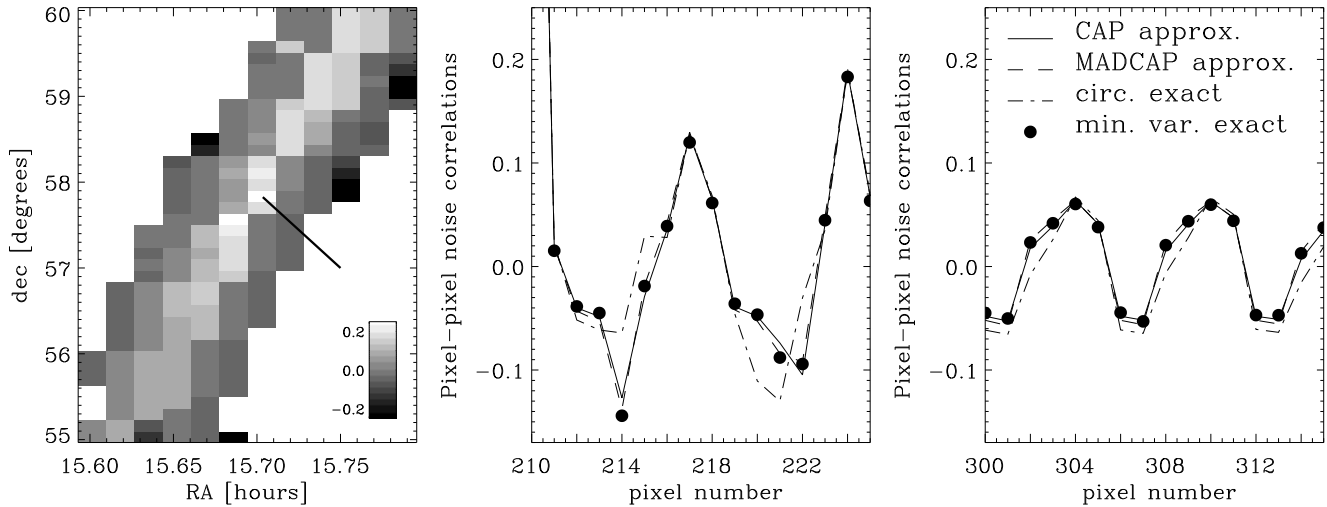}
\vskip -2truecm
\caption{ An example of pixel domain noise correlations estimated for
a single segment of \maximai\ data set. In this figure, we show
correlations of a single selected pixel with its neighbors. This pixel
is marked in left panel with an arrow. The color scale shows the
correlation level relative to the RMS value of the noise in the
selected pixel. The scanning direction was from the left bottom corner
to the top right, leaving a smudge of strong correlations in pixel
domain. Pixels are numbered row by row from the left to the right and from the bottom
to the top as shown in the left panel. 
Note a strongly distorted aspect ratio of the figure due to a
scan elongation in a declination direction.  These correlations as a
function of a pixel number are also displayed in the middle and right
panels. They show the relative correlations of the selected pixel with
its closer (middle) and more distant neighbors (right panel). 
The chosen pixel has a number
210.  Different lines correspond to different map-making methods used
for the correlation estimations. Filled circles show results of the exact
minimum variance, solid line - approximate circulant (CAP), dashed - approximate
MADCAP and dot-dashed - exact circulant methods.  }
\label{mapmakes}
\end{figure*}
Note that the operation count is dominated by the sparse inverse
computation ($n_s > n_{pix}$). The operation count can be reduced if
$\bm{N_t}$ and $\bm{N_{Ct}}$ are assumed to be band diagonal. However,
even then, for experiments like \maxima\ the operation count remains
\order{n_s^2}, usually with a large prefactor. Hence this approach
is far from competitive with those discussed above, although it can be
comparable for very short correlation length data. The memory required
again scales as \order{n_{pix}n_s}.

A possible approximation, which we advocate hereafter, is to neglect
the sparse correction completely in the expression for the pixel-pixel
noise correlation matrix. Clearly the operation count drops to \order{n_s^2}
or \order{n_{pix}^3} -- whichever is larger -- and is usually 
\order{n_{pix}^3} especially if $\bm{N_{C t}}^{ -1}$ is assumed to be band
diagonal. In this case the method's computational requirements are
comparable with those of the MADCAP algorithm~\cite{Borrill1999mad}. In
the following we refer to this approach as the circulant
approximation (CAP). This approximation is clearly very similar to the
MADCAP approach discussed above, although it is strictly exact for a
diagonal noise correlation matrix as well as for an infinitely long
time stream.

\subsection{Comparison and assessment}

\label{assesment:para}

\begin{figure*}
\vskip -4truecm
\leavevmode\epsfxsize=18.cm\epsfbox{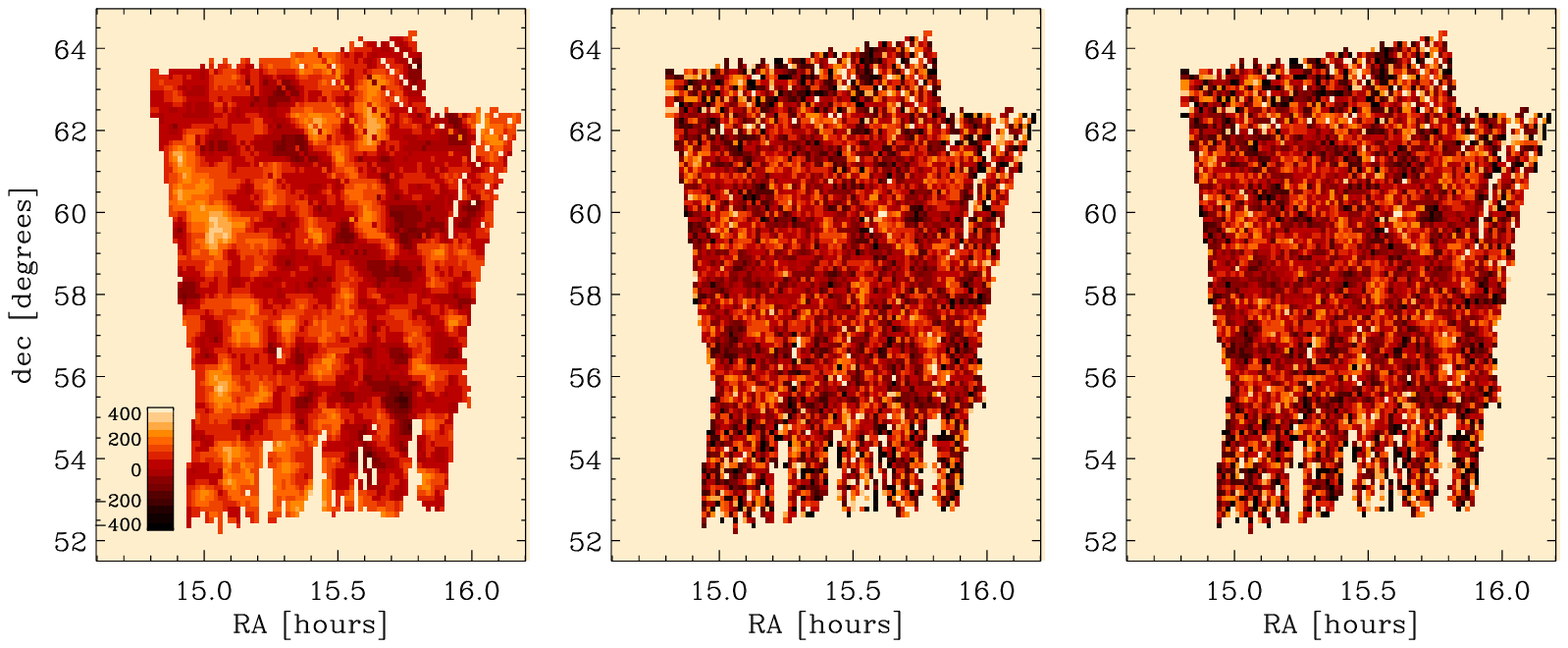}
\vskip -2truecm
\caption{Simulated \maximai-like maps of the CMB anisotropy. Left
panel shows the simulated map used for creating the time stream data
of a single detector of a \maxima-like experiment. The time stream was
subsequently used to recover the map of the sky applying the
approximate CAP (middle) and MADCAP algorithms.  }
\label{simaps}
\end{figure*}

\begin{figure*}
\vskip -4truecm
\leavevmode\epsfxsize=18.cm\epsfbox{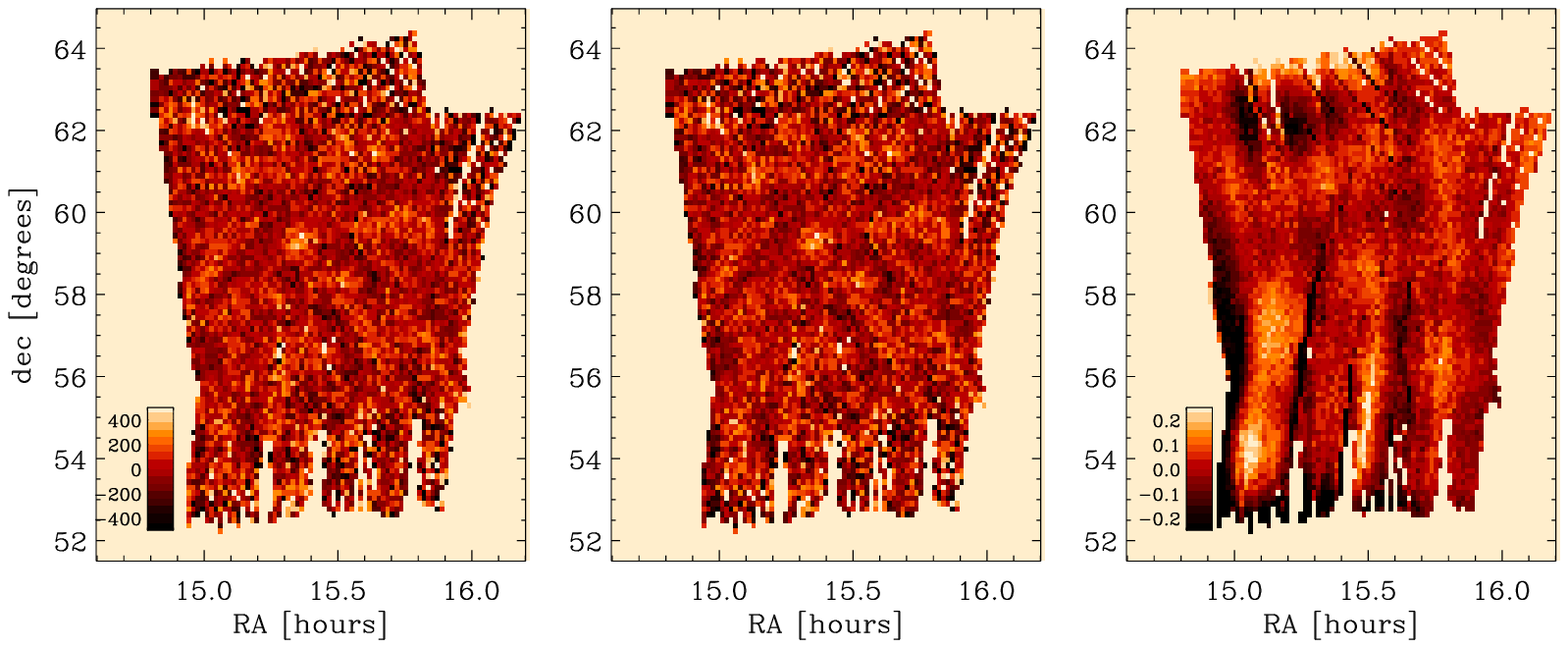}
\vskip -2truecm
\caption{Comparison of two map-making algorithms applied to the real
\maximai\ data of a single photometer. The left panel shows the map
computed using the approximate circulant (CAP) approach. The middle panel
shows the map recovered using MADCAP approach. The difference of both
divided by an estimated rms of the pixel noise is shown in a right
panel. Note that the color stretch limits are from -0.25 to 0.25 in
the right panel.}
\label{rmaps}
\end{figure*}

Although the various map-making methods outlined above are
algebraically similar, being derived from the same underlying
equations, their implementations and generalized extensions to fully
realistic time streams differ considerably.

The approximate methods are of particular interest, because they achieve the
speed of Fast Fourier transforms. If that is attained 
without sacrificing the necessary precision is to be tested and may depend
on 
the particular case at hand.  The important parameters are the noise
correlation length ($\lambda_c$) and the time stream length ($n_s$).
If these numbers are comparable the differences between the methods
can be significant, but they should disappear whenever $\lambda_c \ll
n_s$ and `edge' effects are unimportant.  Given the presence of a
so-called $1/f$ noise component in the time stream, the assumption that
$\lambda_c$ is independent of the time stream length may seem to be plainly
wrong. Instead the correlations should be present on time scales
comparable with the time stream length.  However, a low frequency cutoff
 (\eg, AC high passing filter in the \maxima\ experiment) usually
limits the correlation length independent of the time stream length.

Here we address and illustrate those issues using the \maximai\
experiment as an example.  For this experiment the average length of
the time stream segments is around $50,000$ samples, with some segments as
short as $35,000$ samples. The very gradual decay of the correlations
with time makes it difficult to determine the correlation length
precisely. However, we have found that the dependence of the resulting
map on the value of $\lambda_c$ assumed vanishes before $\lambda_c$
reaches $10,000$ samples, and have therefore used $\lambda_c = 10,000$
in all computations.  Hence, for \maximai, $\lambda_c$ is less than
but not negligible to $n_s$, and prior to taking advantage of the
speed of the approximate methods we need to demonstrate their
applicability.

Note that comparing the different methods is not entirely
straightforward. Assumptions about the band-diagonality of either the
time domain noise correlation matrix or its inverse, as required by
the different methods, are clearly not equivalent. The relation
between the assumed bandwidths of these matrices is also not obvious.
It may seem that a fair comparison would be to take the bandwidths to
be as large as possible in all methods. However, we then run into
the problem of numerical inaccuracies unavoidably present in the
calculation of the noise correlation matrix (or its inverse),
especially at large time lags, which can result in a non-positive
definite (unphysical) noise correlation matrix in pixel
domain. Although power spectrum windowing (see \eqns~\ref{nctfull},~\ref{invnctfull}) can
be used to alleviate numerical problems of this sort, this may also
have different consequences for each of the methods.

This point is particularly conspicuous when comparisons are made with
the exact circulant method. We find that this approach is very
sensitive to both the choice of $\lambda_c$ and the windowing
technique. For this reason we do not quote precise numbers for the
comparison of this method with the others. By adjusting the free
parameters of the method we can reproduce other methods' results quite
accurately, and within the expected statistical uncertainty of the maps.

In the case of the other comparisons we have applied a Gaussian window
function (\eqns~\ref{nctfull},~\ref{invnctfull}) while computing the noise power spectrum and kept similar
correlation lengths for all three methods. In any case, the numbers
quoted below should be looked at as indicative rather than as the best
case possible.

We choose to compare maps and noise correlation matrices directly in
pixel domain, notwithstanding the fact that the approximations are
really applied when the inverse of the noise correlation matrix in the
time domain is computed (see Fig.~\ref{tnoise:fig}). It is the quality of the estimate of the map
and its noise correlation matrix which matter for any subsequent
analysis. Such an approach also seems to be more general than assessing
the quality of a map through the application of a specific statistical
tool.

A sample of results is shown in
Fig.~\ref{mapmakes}. The middle and right panels show two regimes of
pixel-domain noise correlations as estimated using different methods. The
left panel shows the correlations of a selected pixel with all the
others as projected on the sky. These were computed for a single $\sim
40,000$-sample segment of the CMB1 scan covering an area of $\sim 6$
square degrees on the sky, corresponding to $\sim 400$ square 8arcm
pixels.  Neither filtering nor marginalization has been applied to
the time ordered data.  The agreement between results from the
different methods is generally quite good.  More quantitatively, $\sim
98$\% of the matrix elements show relative differences less than $20$\% when
computed using the exact minimum variance approach and the MADCAP approximation, and $\sim 50$\%
show differences less than $5$\%. Similar numbers are found for the
comparison of this exact method and the CAP approximation.
Usually, both approximate methods tend to overestimate high positive
correlations. The MADCAP approach also seems to underestimate the
amplitude of negative correlations, yet frequently recovers weak
correlations more precisely than the other approximation. The relative
differences of the maps are of the order of $10$\%
(Fig.~\ref{simaps}), which, on average, amounts to no more than
$10$\% of the rms noise level predicted for a given pixel
(Fig.~\ref{rmaps}).

To test if the discrepancies are due to numerical errors, rather than
differences in the algorithms, we use the exact method with an
explicitly circulant noise correlation matrix, and compare the result
with that calculated using the approximate circulant approach. In the
absence of numerical inaccuracies both results should be identical.
We find that the differences are predominantly at the $2-3$\% level.
We obtain similar error estimates when analyzing a diagonal noise
correlation matrix using all three methods.

We can also ask if there is any systematic error incurred as a
consequence of the approximations which bias the results of a
particular approach. One way to address this issue with simulated maps
is to check whether the actual noise in an estimated map is properly
described by a concurrently estimated noise correlation matrix. This
is clearly the case for the exact methods, which are derived to obey
such a requirement.  To test the approximate approaches we define
$\bm{m_{sky}}$ as the true noiseless map of the sky used for a
simulation.  $\bm{m_{c}}$, $\bm{N_c}$, and $\bm{m_{m}}$, $\bm{N_m}$
are maps and pixel-pixel noise correlation matrices as estimated by
the CAP and MADCAP approximate methods respectively. For each of the
noise matrices we introduce a matrix $\bm{N_i}^{-1/2}$ such that
$\bm{N_i}^{-1/2}\l(\bm{N_i}^{-1/2}\r)^T=\bm{N_i}^{-1}$.  Consequently the
variable $\bm{y_i}$, defined as
\begin{equation}
\bm{y_i}\equiv \bm{N_i}^{-1/2}\l(\bm{m_i}-\bm{m_{sky}}\r),\ \ \ \bm{i=c,m},
\end{equation}
should be a vector of uncorrelated, Gaussian variables with a
dispersion equal to unity if no systematic problem is introduced by an
approximation. This can be tested using, \eg, the Kolmogorov-Smirnov
test (see also Sect.~\ref{constest:para}). Such a test is comfortably passed by both methods for noise in
the pixel domain ranging from $\sim 100\mu$K per 8 arcminute pixel (the
\maximai\ single channel level) down to $\sim 60\mu$K (the \maximai\
4-channel combined level).

The differences between the results of the various methods, as
summarized above, are found to be rather small, and indicate that all
the methods may be equivalent in practice.  Although the problem is
difficult to tackle in a more quantitative and general way, it can be
addressed from the point of view of specific statistics derived from
the maps. An example of such a test, comparing the power spectrum
statistic, is shown in Fig.~\ref{cl_comp}. In this case the power
spectra of the \maximai\ maps computed using both map-making methods
show very good agreement.

\begin{figure}
\vskip -0.25truecm
\leavevmode\epsfxsize=8.cm\epsfbox{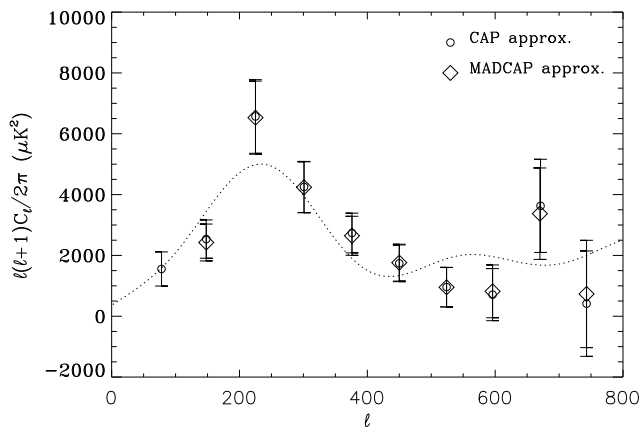}
\vskip 0.0truecm
\caption{
Comparison of power spectra computed for the maps made with two
different map-making algorithms as shown in two left panels of Fig.~{\ref{rmaps}}.
The data are those of a single \maximai\ detector. Circles show the
results for the map made using approximate circulant (CAP) algorithm, and
diamonds show the results for the map made using the MADCAP approximation.
}
\label{cl_comp}
\end{figure}

In practise, the approximate methods have an advantage over the exact
methods of being easily implementable. In fact, both can be
straightforwardly applied using the MADCAP package~\cite{Madcap}.
The differences in the
results provide some insight into the sensitivity of the results to
the treatment of time domain noise correlations.  More to the point,
the result of any statistical test applied to the map, which depends
on the choice of the approximate method used during map-making, should
be treated with suspicion. We have demonstrated that for the \maximai\
data analysis and, in particular, for power spectrum estimation, the
differences between the methods are negligible.

In summary, all four map-making methods are broadly consistent at the
noise level of the \maximai\ maps. This concordance is only expected
to improve if longer time stream segments are considered. Numerical
errors can readily be kept under control even with as many as 
\order{10^5-10^6} time samples, at least for \maxima-like data sets, although
further tests may be needed for low noise cases. Both approximate
methods are easy to implement, with their speed being limited by the
pixel domain noise correlation matrix inversion.
The exact minimum variance method 
provides a useful cross-check on both the validity of the
approximations and numerical error propagation, and, with further
improvements may be able to achieve the computational speed of the
approximate approaches.

We note again that all of this assumes a continuous time stream with
stationary noise resulting from application of the gap filling
algorithm; the restoration of the Toeplitz character of the noise
correlations is crucial for the feasibility of the exact minimum
variance approach, as well for as the accuracy of the approximate
methods.

\section{Map-making: amendments}

\label{amendments}

So far we have been assuming an `optimistic' model for the signal in
the time stream, $\bm{d_t}$ (\eqn~\ref{simpletimestream}).  Commonly,
unwanted contributions (\eg, the overall average temperature, primary
mirror- or gondola- synchronous noise) are present in the data
(\eqn~\ref{simtstream}) and have to be dealt with if a reliable result
is to be obtained. 

Let us consider a case of an arbitrary time domain contribution of known temporal
behavior. We take these to be given by a set of $m$ template vectors $\bm{\tau_t^{\l(i\r)}}$,
of unknown amplitudes, $x^{\l(i\r)}$, so the resulting time stream equation is
\begin{equation}
\bm{d_t}=\bm{A} \bm{m_p}+\bm{n_t}+\sum_{i=1}^{m} x^{\l(i\r)} \bm{\tau_t^{\l(i\r)}}.
\label{tstream1:eq}
\end{equation}
The additional number $x^{\l(i\r)}$ can be formally treated as an extra (fictitious) pixel
`observed' with a `pointing matrix' as given by
$\bm{\tau_t^{\l(i\r)}}$. These additional numbers can be estimated during the course of a 
map-making procedure and then marginalized over.
Alternatively, the entire extra term, $\sum_i x^{\l(i\r)}\bm{\tau_t^{\l(i\r)}}$, can be 
thought of as a part of the noise term and directly marginalized over in the course of
map-making.
Clearly these two approaches are just different implementations of the same idea, yet,
depending on the particular problem at hand, one or the other may be more appropriate.
In the following, we refer to them
as the {\em extra pixels} and {\em marginalization} methods, respectively, and discuss them and some of their
applications.

\subsection{Extra pixels.}

\label{extrapixels:para}

Let us define a vector $\bm{x_q}$ as, 
\begin{equation}
\bm{x_q}^T\equiv\l[x^{\l(1\r)}, \cdots, x^{\l(m\r)}\r],
\end{equation}
and a matrix $\bm{B}$ such that,
\begin{equation}
\bm{B}\equiv\l[\bm{\tau_t^{\l(1\r)}},\cdots,\bm{\tau_t^{\l(m\r)}}\r].
\label{Bdef:eq}
\end{equation}
We can now rewrite \eqn~(\ref{tstream1:eq}) as,
\begin{equation}
\bm{d_t}=\bm{A}\bm{m_p} + \bm{B}\bm{x_q} + \bm{n_t}.
 \label{tstream2:eq}
\end{equation}
From this it is clear that the extra time stream contribution can be
thought of as
a set of extra (fictitious) pixels (\bm{x_q}) `observed' with a pointing matrix given by $\bm{B}$.
If we, furthermore, recast this equation as,
\begin{equation}
\bm{d_t}=\l[\bm{A},\bm{B}\r]\l[\begin{array}{c}\bm{m_p}\\ \bm{x_q}\end{array}\r]+\bm{n_t}
\equiv {\gA}  {\gm}  +\bm{n_t}, \label{realtimestream}
\end{equation}
we find that it is formally identical to 
\eqn~(\ref{simpletimestream}). Here, we have introduced a generalized
pointing matrix, $\gA,$ and a generalized map, $\gm$. Those correspond
to an `ordinary' map and a pointing matrix, but now extended to
incorporate also the extra pixels.  The map-making methods discussed
above can all be applied in the present case, with all their caveats and
strengths.  The appropriate equations preserve the form of 
\eqns~(\ref{mapmaking}) \&\ (~\ref{noisemap}) but the pointing matrix, $\bm{A}$, the map,
$\bm{m_p}$, and the pixel domain noise correlation matrix, $\bm{N_p}$,
need to be replaced by their generalized counterparts, \ie, $\gA,$ $\gm$
and $\gN_{p}$ respectively.  As a result, the map-making procedure
provides an estimate of a generalized map and a generalized noise
correlation matrix in pixel domain.  

In addition to the minimum-variance considerations described above, the
formalism so far described can also be thought of as describing the
likelihood function for the data:
the distribution of the underlying (generalized) map is Gaussian with
mean and variance as given by \eqns~(\ref{minvarmap}) \&\
(\ref{minvarnoise}) (or approximations to these). Thus, the generalized
map is a simultaneous estimate of both the real map and the `extra'
pixels. Knowledge of the amplitudes of these extra pixels are often
useful for diagnostic purposes. However, we are in the end interested in
the real map itself; hence we wish to marginalize over the
$\bm{x_q}$. With a uniform prior, we find the usual Gaussian results:
\begin{itemize}
\item the marginalized map is just the generalized map with the extra
  pixels removed; and
\item the marginalized noise matrix is just the generalized noise matrix
  with the rows and columns corresponding to the extra pixels removed.
\end{itemize}

Clearly, only well-understood features of the time stream, for which the pointing
operator $\bm{B}$ is known, can be accounted for using this approach.
Moreover, if the extra
component is indistinguishable from the CMB temperature map, \ie, is sky-stationary,
then the clean separation of the map into CMB and non-CMB components is impossible.
That is manifested as a singularity of the $\gA^T \bm{M} \gA$ (see
\eqns~\ref{mapmaking} \& \ref{noisemap}) matrix. In some cases, if such
a singular pixel-domain mode is determined, it can be accounted for on
subsequent stages of the analysis (see Sect.~\ref{singularities:para}).
In general the discussed framework is quite universal and can be applied
successfully in a variety of circumstances of practical interest.
Specific examples of its applications are time stream gaps, relative offsets between separate segments of the
map, and primary mirror synchronous effects. All those were found of
importance for the \maximai\ data analysis and we discuss them in detail below.

\paragraph{Time-stream gaps, offsets and temporal frequencies.}

Perhaps the most straightforward applications of this method are to
time stream gaps, time-stream segments offsets, and unwanted temporal 
frequencies. 

The direct application of the extra pixel method to the unprocessed raw
time stream in order to properly marginalize over the unknown content of
the gaps would require introducing as many extra pixels (parameters
$x_i$ in \eqn~\ref{tstream2:eq}) as time samples in the gaps.  Not only
is this a computationally significant extension, but also a source of a
multitude of possible numerical problems and singularities.  With the
gap-filling procedure as described earlier, one extra $x$ parameter (the
gap pixel) and a single template, $\bm{\tau_t}$, such as,
$$\bm{\tau_t}\l(i\r)=\l\{\begin{array}{l l} \dsp{1,} & \dsp{i\,\in\,
    gaps;}\\ \dsp{0,} & \dsp{\hbox{otherwise;}}\end{array}\r.$$
suffices
to account for all of the gaps of a single segment.  That is precisely the
gap pixel approach briefly mentioned in Sect.~\ref{minvarmaps:para}.

If a map of more than a single segment is to be produced
from a single map-making procedure (an approach adopted in MADCAP), care
must be taken with regard to unknown relative offsets between different
segments. In total power experiments the actual offsets are spurious and
contain no information about the sky. In the parlance of the extra pixel
method, the offsets can be considered as amplitudes, $x^{\l(I\r)}$, of a set of
time-domain templates, \bm{\tau_t^{\l(I\r)}}, defined for each segment $\bm{I}$ as
$$\bm{\tau_t^{\l(I\r)}}\l(i\r)=\l\{\begin{array}{l l} \dsp{1,} & \dsp{i\,\in\,
    I;}\\ \dsp{0,} & \dsp{\hbox{otherwise;}}\end{array}\r.$$
and
therefore they can be straightforwardly incorporated within the framework
of the extra pixel method and the map-making
formalism~\cite{Borrill1999mad}.

Similarly unwanted temporal frequencies can be described with one extra
parameter, their common amplitude, if they are first filtered out from
the time stream, and then replaced by a pure Gaussian noise realization
with the noise power spectrum as estimated earlier.

\paragraph{Primary mirror synchronous signal.}

\begin{figure*}[t]
\vskip -4truecm
\leavevmode{\epsfxsize=18.cm\epsfbox{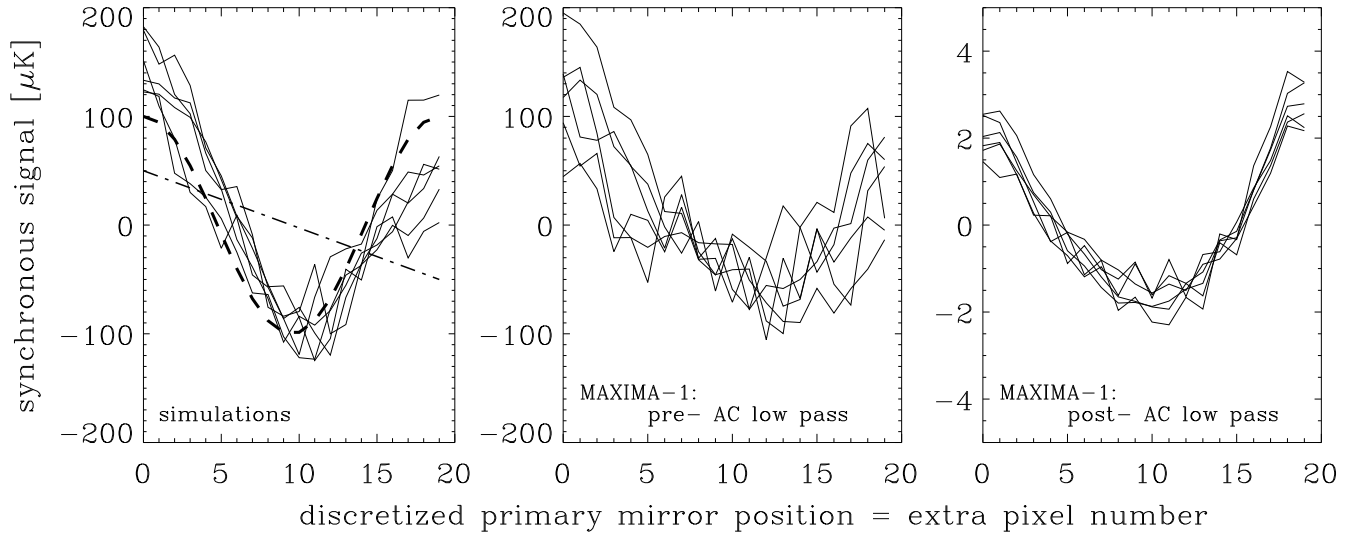}}
\vskip -2truecm
\caption{The primary mirror synchronous signal recovered from the
simulation (left panel) and the real \maximai\ data as a function of
the discretized position of the primary mirror with respect to the gondola.
In each of the panels, the recovered signal for each of the six segments of the CMB1
scan is shown with thin solid line. In the left panel, the thick
dashed line shows the primary mirror synchronous signal as used for
the simulation. The apparent asymmetry of the recovered signal in contrast
with the symmetry of the function used for simulation is due to a residual
CMB dipole contribution shown with a thin dot-dashed line. A very
similar, slightly asymmetric, shape of the functional dependence is found
for the signal recovered from the real data as shown in the middle panel.
That panel shows the estimated amplitude of the signal synchronous with
the mirror position, which was added to the time stream data prior to
the AC low-pass filtering and therefore also includes the sky signal,
such as the CMB dipole. The rightmost panel shows the estimated primary mirror
synchronous signal added to the time stream after AC low pass
filtering, and therefore originating in the instrumental
electronics. Note the very good stability of that contribution with
time and its negligible amplitude, as compared to the signal
shown in the middle panel, as well as to the expected CMB anisotropy.
The noise level corresponding to different mirror positions is
strongly correlated and for that reason not show in the figure. A
typical level is $\sim 40\mu$K and $\sim 1\mu$K for the middle 
and left panel.
}
\label{choptemps}
\end{figure*}

Periodic motion of the primary mirror and the gondola can potentially
become a source of an extra parasitic contribution to the total signal measured by a
\maxima-like experiment. Due to its origin, such a contribution is
likely be a single-valued function of the position of either the primary
mirror or the gondola and therefore can be modeled using the extra
pixel method discussed above.

Only the primary mirror synchronous effect has been found to be
important for the actual \maximai\ data and required this treatment.  
Below we describe this case in some detail as an example.

Assuming that the primary mirror synchronous  
contribution is a slowly varying function
of the mirror position, we can characterize it using a discrete set of
amplitudes (\ie, an extra pixel `map'), each of which describes the
magnitude of the parasitic signal corresponding to a narrow range of mirror
positions (\ie, an extra pixel).  The extra pointing matrix $\bm{B}$
assigns then the time samples to these discretized primary mirror positions.
The presence of the instrumental filters introduces an additional
complication. 
They induce phase shifts in the time stream and therefore modify the pointing matrix in a way depending on the
precise location that the synchronous signal appeared in the on-board
electronics chain (which we do not know {\em a priori}).  In principle, there are four possible choices for
the correct pointing matrix of the primary mirror synchronous signal.
However, the effective, total, phase shift, caused by the instrumental filters, is strongly dominated
by the AC low pass filter, leading, therefore, to only two
significantly different choices for the combined pointing matrix of this synchronous effect,
${\cal B}$.
We choose those to be,
\begin{eqnarray}
{\cal B}&=&\bm{B},\nonumber\\
\mbox{or, \ \ } {\cal B}&=&\bm{F_{bolo}}^{-1} \bm{F_{low}}^{-1} \bm{F_{high}}^{-1} \bm{B}.\nonumber
\end{eqnarray}
Here we use $\bm{B}$ to denote the discretized mirror position and
denote instrumental filters as in Sect.~\ref{timestream:para}.
The first choice above corresponds to the case in which the synchronous signal is added prior to
AC low pass filtering, \eg, a mirror related modulation,
and the second one to the case in which the extra contribution arises later on.

A number of extra pixels depends on a particular problem at hand.
In the \maximai\ case  we use a separate set of extra pixels, consisting of $20$ to
$50$ discretized mirror positions, for each of the segments.  The relative offset of the primary
mirror signal and the CMB map is arbitrary, therefore we constrain the
value assigned to the first extra pixel (corresponding to the leftmost
position of the mirror) to be zero.  That breaks the degeneracy between
absolute offsets for both components and avoids a singularity in the
generalized noise correlation matrix in pixel domain.

For \maximai\ we find no other singularity caused by the inclusion of extra
pixels describing the primary mirror synchronous signal in
the map-making problem.
This is due to the carefully designed \maximai\ scanning strategy;
the typical time scale for variation of the extra contribution 
is significantly longer than the time needed for the instrument
antenna to cross the pixel on the sky, and the majority of real
sky pixels are revisited many times, with the primary
mirror position different at each visit.

By using a single set of extra pixels per segment, we make an implicit
assumption about the stationarity of the underlying synchronous signal
on the time scale of the length of the segment.  Such an assumption needs
to be tested {\em a posteriori}. Recovered primary-mirror-synchronous signals
are shown in Fig.~\ref{choptemps}. These results are for simulations
with the synchronous signal explicitly assumed to be strictly
stationary, and for the actual \maximai\ data.  In the latter case, the
results for different, but adjacent, segments of the same detector are
indeed consistent within the error bars, suggesting that the
assumption of stationarity is satisfied.

\begin{figure*}[t]
\vskip -5 truecm
\leavevmode{\epsfxsize=18.cm\epsfbox{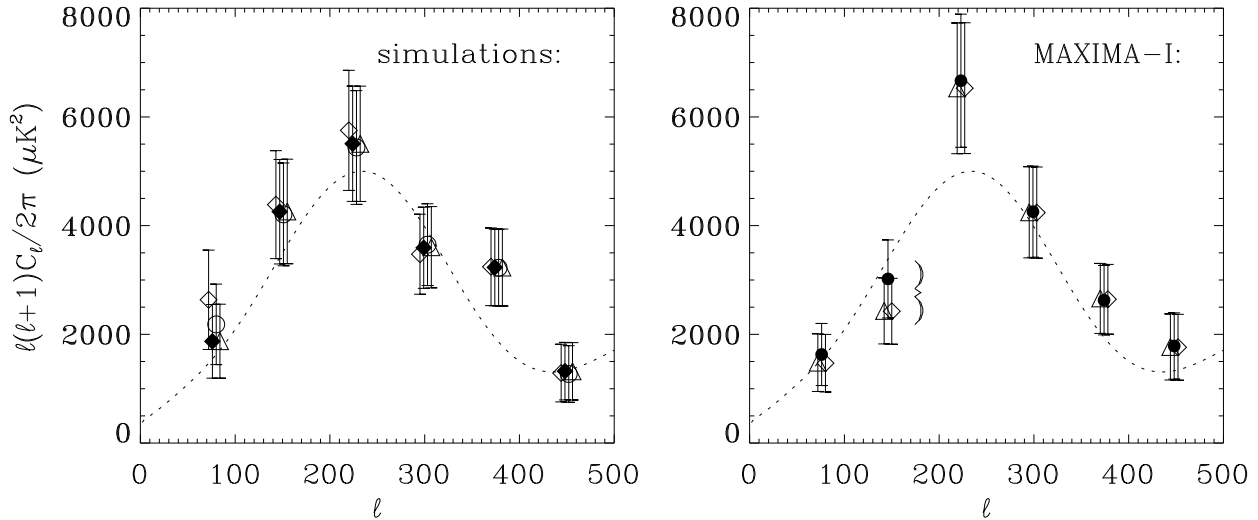}}
\vskip -1truecm
\caption{Power spectra of the CMB anisotropies recovered from the
  simulations (left panel) and the \maximai\ single detector map
  (right). In the left panel all shown spectra are computed assuming the
  same CMB sky and noise realization and a \maximai\ like observation.
  We demonstrate here the performance of the extra pixel method in a
  case of primary synchronous signal. The displayed points correspond to
  four cases: with a primary synchronous signal absent in the time
  stream data and either not accounted for during map making (triangles)
  or accounted for (open circles) using the extra pixel method on the
  map-making stage.  The remaining two cases, with a primary mirror
  signal present in the data are shown with open diamonds (no extra
  pixel method applied) and filled diamonds computed for the map
  produced with the extra pixel method.  Right panel shows results of
  tests of the extra pixels method applied directly to the data. Spectra
  shown with open symbols have been computed using different definitions
  of the extra pixels: diamonds are for the standard case with
  synchronous signal depending only on the position of the antenna, and
  triangles are for the case when also a dependence on a direction of
  the primary mirror motion (left or right) is allowed.  Filled circles
  depict the spectrum computed for a map with no extra pixel method
  applied during map making. Clearly, only the power in the bin centered at $\sim 150$
  has been affected by the primary mirror synchronous signal.
 }
\label{chopcl}
\end{figure*}

The differences between recovered signals can be the result of 
actual differences of the instrumental signal, or of the sky signal,
which can have a component synchronous with the mirror position, or
because of instrumental noise.
We find that for the \maximai\ experiment,
a part of the dipole signal is subtracted from the sky map together
with the primary mirror synchronous signal. 
This effect is mainly due to the
monotonic gradient-like morphology of the dipole within the boundaries
of the $\sim 10^\circ\times10^\circ$ patch observed by \maximai.  For all
multipole modes which vary significantly across that patch, \ie, for
$\ell \simgt 20$, such an effect is expected to be unimportant. We show
effects of the subtraction of the primary mirror synchronous signal on a
power spectrum estimation in Fig.~\ref{chopcl}.

\subsection{Marginalization.}

\label{alternative:para}

In the extra pixel method, we first determine the generalized map and
noise matrix, and then marginalize over the unwanted pixels.
In the case of the marginalization approach, we first marginalize over
the unwanted temporal modes, and then pass directly to the
marginalized map and noise matrix.  It is based on the idea that the
undetermined amplitude can be treated as a random variable with a prior
probability with dispersion $\sigma_x^2$~\cite{BJK1998}. Again, we start
with the more complicated time stream of \eqn~(\ref{tstream1:eq}), but for compactness
allow only for a single template, $\bm{\tau_t}$.
Now, we
assign the unknown amplitude, $x$, a prior probability density with variance
$\langle x^2 \rangle = \sigma_x^2$. Then, we
marginalize over this unknown amplitude right away, before making the
map.  We then recover a distribution for the time stream in the form of a
Gaussian with an effective time stream noise matrix given by
(hereafter $\otimes$ denotes a tensor product),
\begin{equation}
\bm{N'_t}=\bm{N_t}+\sigma_x^2 \bm{\tau_t}\otimes\bm{\tau_t}^T.
\end{equation}
That is, we can consider both the Gaussian noise ($\bm{n_t}$) and the
template ($x\bm{\tau_t}$) together as a noise-like term with this
correlation matrix. The more complicated equation
(\ref{tstream1:eq}) can be then recast in the familiar form of
\eqn~(\ref{simpletimestream}).  Solving for the maximum
likelihood map gives expressions in the pixel domain mirroring that of
\eqn (\ref{mapmaking}), with the noise correlation matrix in time
domain replaced now by $\bm{N'_t}$. Marginalizing over the amplitude,
$x$, while taking the limit $\sigma_x^2\to\infty$, is equivalent to
marginalization with a uniform prior, as considered
above~\cite{BJK1998}.  This is tractable because we can simplify the
calculation of $\bm{N'_t}^{-1}$ using the Sherman-Morrison-Woodbury
formula~\cite{Press1992}, in this context given by
\begin{eqnarray}
  {\bm{N'_t}}^{-1}=\bm{N_t}^{-1}&-&\frac{\l(\iNtt \bm{\tau_t}\r)\otimes
    \l(\iNtt \bm{\tau_t}\r)^T}{\sigma_x^{-2}+\bm{\tau_t}^T
    \iNtt\bm{\tau_t}}\nonumber\\ 
  \myatop{_{\sigma_x^2 \rightarrow \infty}}{\longrightarrow}
   \iNtt&-&\frac{\l(\iNtt \bm{\tau_t}\r)\otimes \l(\iNtt
    \bm{\tau_t}\r)^T}{\bm{\tau_t}^T \iNtt\bm{\tau_t}}. 
  \label{margnoise}
\end{eqnarray}
As expected we have,
\begin{equation}
\bm{N'_t}^{-1}\bm{\tau_t}=0.\label{zeroeigen:eq}
\end{equation}
Finally, note that the usual map-making formulas in a minimum variance case
(\eqns~\ref{minvarmap} \&\ \ref{minvarnoise}) require
only the inverse noise correlation matrix, $\bm{N_t}^{-1}$, and
remain unchanged if $\bm{N_t}^{-1}$ is
replaced by $\bm{N'_t}^{-1}$. Because the correction to $\bm{N_t}^{-1}$
is additive, we can also think of this as a correction to the inverse
pixel noise:
\begin{equation}\label{eq:noiseprime}
  \bm{N'_p}^{-1} \to \bm{N_p}^{-1} - \frac{\l(\bm{A}^T\iNtt
    \bm{\tau_t}\r)\otimes \l(\bm{A}^T\iNtt \bm{\tau_t}\r)^T}
  {\bm{\tau_t}^T \iNtt\bm{\tau_t}}.
\end{equation}

One might suspect that a weakness of the marginalization method would be the
computational overhead involved in computations of the extra term in
\eqn~(\ref{margnoise}).  However, the products of $\iNtt$ matrix and
a template vector $\bm{\tau_t}$ requires only \order{n_s \log n_s}
operation if $\iNtt$ is Toeplitz, or \order{n_s \lambda_{band}} if 
it is (approximately) band-diagonal with a band-width, $\lambda_{band}$.
Also, recalling
that what we need in \eqn~(\ref{eq:noiseprime}) is
$\bm{A}^T{\bm{N'_t}}^{-1}\bm{A}$ rather than $\bm{N'_t}^{-1}$ itself, we
can cut the number of necessary operations by performing the products
from outside inwards. That can be done in either \order{n_s} or \order{n_{pix}^2}
floating point operations, whichever is larger. So the final scaling for
any single template is given either by \order{n_s \log n_s} or
\order{n_{pix}^2}.

If more than one template is required than \eqn~(\ref{margnoise}) needs to be generalized and reads,
\begin{equation}
\bm{N'_t}^{-1}=\bm{N_t}^{-1}-\l(\bm{N_t}^{-1}\bm{B}\r)\l[\bm{B}^T\bm{N_t}^{-1}\bm{B}\r]^{-1}\l(\bm{N_t}^{-1}\bm{B}\r)^T
\label{multitemp:eq}
\end{equation}
where $\bm{B}$ is as defined in the previous Section (\eqn~\ref{Bdef:eq}), and the expression in the square brackets
is a square matrix of the size given by a total number of time-domain
templates. We have assumed that the template amplitudes are
uncorrelated, \ie, $\l\langle x^{\l(i\r)}x^{\l(j\r)}\r\rangle=0$ if
$i\ne j$.
In this case again (\cf,~\eqn~\ref{zeroeigen:eq}),
\begin{equation}
\bm{N'_t}^{-1}\bm{B}=0,\label{zerotemp:eq}
\end{equation}
and, hence, also for each template, $\bm{\tau_t^{\l(i\r)}}$, (\cf,~\eqn~\ref{Bdef:eq}),
\begin{equation}
\bm{N'_t}^{-1}\bm{\tau_t^{\l(i\r)}}=0.\label{zeromultieigen:eq}
\end{equation}
This approach is fully general and can be used in all the cases already
mentioned in the context of the extra pixel method --- both methods give
identical results. The advantage of the extra pixel method is that it
not only marginalizes over the unwanted effect, but also computes its
individual maximum-likelihood estimate.  That can be useful for {\it a
  posteriori} tests, if some prior assumptions have been made to extract
the effect.

Note also that we have constructed a matrix we call
$\bm{N'_t}^{-1}$. However, in the limit $\sigma_x^2 \rightarrow \infty$,
$\bm{N'_t}$ itself no longer exists: by construction it has an infinite
eigenvalue corresponding to the eigenvector $\bm{\tau_t}$. 
(There are also pathological cases where $\bm{N'_t}^{-1}$ can be inverted 
because of numerical effects.)
Nonetheless,
the pixel-domain noise matrix $\bm{N'_p}$ ($\equiv(\bm{A}^T{\bm{N'_t}}^{-1}\bm{A})^{-1}$) may still
exist due to mixing of the modes via the operator $\bm{A}$.
In fact the resulting $\bm{N'_p}^{-1}$ is singular only if 
$\bm{\tau_t}=\bm{A}\bm{v_p}$, where $\bm{v_p}$ is a pixel domain template.
%is defined in a pixel domain.
Then from \eqn~(\ref{eq:noiseprime}) it follows that (cf.~\eqn~\ref{zeroeigen:eq}),
\begin{equation}
\bm{N'_p}^{-1}\bm{v_p}=0.
\label{mapsingmode:eq}
\end{equation}
In cases when $\bm{N'_p}$ exists the `extra pixel' approach will work as well. However, the
marginalization procedure would work even if
$\bm{A}^T{\bm{N'_t}}^{-1}\bm{A}$ is singular as we discuss that in
Sect.~\ref{singularities:para}.

Below we elaborate on a particular application of the marginalization
method to the removal of specific frequency modes from the time stream.

\label{marginalization:para}
\setcounter{paragraph}{0}
\paragraph{Frequency band marginalization.}
Here we will use the method of the explicit marginalization, outlined
above, to derive a useful concise formula marginalizing over unwanted frequency bands.
In such a case a required set of templates is given in the time domain as,
\begin{equation}
\bm{\tau}_{\l(m\r)}\l(j\r)\equiv \cos\l(\frac{2\pi i_0\l(j+m\r)}{n_s}\r).
\label{freqtemp:eqn}
\end{equation}
Here, $i_0$ corresponds to the frequency $f_0=i_0/ n_s\Delta$,
which is to be marginalized over, $j$ is a time variable, and
$m=0,...,[n_s/i_0]$ determines the overall phase shift and also
numbers the template. As usual $n_s$ stands for the length of the
time stream segment under consideration and $\Delta$ is the sampling rate.
We can now insert these templates into \eqn~(\ref{multitemp:eq}) to obtain
a fully general formula for the inverse noise correlation matrix with a frequency
mode, $f_0$, marginalized over. As explained above, a numerical implementation
of such a strict marginalization 
is feasible within a framework of the minimum variance methods.
This expression simplifies significantly if $\iNtt$ is a
circulant matrix, $\iNtt=\bm{N_{Ct}}^{-1}$. Then ${\bm{N'_t}}^{-1}$ is also circulant, 
${\bm{N_t'}}^{-1}=\bm{N'_{Ct}}^{-1}$, and both are uniquely defined by their
power spectra (\cf, \eqn~\ref{nct}). Here, we denote these as ${\cal P}'\l(f\r)$
and ${\cal P}\l(f\r)$. From \eqn~(\ref{multitemp:eq}) we then have
\begin{equation}
{\cal P}'\l(i\r)={\cal P}\l(i\r)-{\cal P}\l(i_0\r)\l(\delta^K\l(i,i_0\r)+\delta^K\l(i,-i_0\r)\r).
\label{powmarg:eqn}
\end{equation}
Here $\delta^K$ is the Kronecker delta, and we have made use of the fact that in Fourier space the
template $\bm{\tau}_{\l(m\r)}$ is represented as,
\begin{eqnarray}
\htaum\l(j\r)&\equiv& \cos\l(\frac{2\pi i_0 m}{ n_s}\r) \l[\delta^K\l(j,i_0\r) +\delta^K\l(j,-i_0\r)\r]\\
&+&\iota \sin\l(\frac{2\pi i_0 m}{ n_s}\r)  \l[\delta^K\l(j,-i_0\r) -\delta^K\l(j,i_0\r)\r],\nonumber
\end{eqnarray}
with $\iota$ denoting an imaginary unit.
From \eqn~(\ref{powmarg:eqn}), we have ${\cal P}'\l(i_0\r)={\cal
P}'\l(-i_0\r)=0$, and therefore, in the case of circulant noise matrices, the extra marginalization term in
\eqn~(\ref{multitemp:eq}) just zeroes the power of the frequency mode
which is to be marginalized over.
In fact, such an answer could have been guessed, by noting that the single-frequency modes are eigen-modes 
of the inverse of the circulant matrix (as well as the circulant matrix itself) and that to introduce a 
Sherman-Morrison-Woodbury-like correction with respect to any of those
modes it is sufficient to set their corresponding eigen-values to zero (\eqn~\ref{zeromultieigen:eq}).

This observation suggests a simple prescription for frequency
marginalization in the case of the circulant approximate map-making algorithm
discussed before.
Because the power spectrum related to the matrix $\bm{N_{Ct}}^{-1}$ is the inverse of the noise power spectrum
(\cf, \eqn~\ref{invnctfull}), it is possible to marginalize over the amplitude of a
frequency mode, $f_0$, by setting the inverse of the noise power spectrum
corresponding to that frequency to zero prior to calculating
$\bm{N_{Ct}}^{-1}$ via \eqn~(\ref{invnctfull}).

The impact of such a procedure on the final map is clear from
\eqn~(\ref{mapmaking}): 
the frequency modes with zeroed power are removed from the map but
that is self-consistently accounted for in the corresponding noise correlations matrix.

Similarly, the MADCAP approach can be modified by adopting the following 
approximation to the inverse noise correlation matrix in time domain
(\cf, \eqn~\ref{FJapprox}),
\begin{equation}
\bm{N'_t}^{-1}\l(i,j\r)\simeq {\bm{N'_{Ct}}}^{-1}\l(i,j\r), \mbox{\ \ if\ \ } |i-j| < n_s/2,
\label{fjmarg}
\end{equation}
and zero otherwise.
The eigenvectors of $\bm{N'_{t}}^{-1}$ are not in fact
identical with those of $\bm{N'_{Ct}}^{-1}$ and, therefore, the former will not usually have 
the eigenvalues equal precisely to zero anymore. The removal of unwanted modes from
the map in this case is therefore only approximate. Again, this is an
effect we have found to be negligible in practise.

One may wish to marginalize over frequency bands which 
are compromised by a periodic parasitic signal (\eg, synchronous
effects). However, in this case this method is less discriminating
than the extra pixel method, making no use of the phase information
usually available.
Marginalization is also useful to minimize the
significance of sampling uncertainty present at low frequency
and leading to errors in the noise estimation procedure (see Sect.~\ref{noisestim:para}). It can also be
applied at the high frequency end, where the precise shape of the
instrumental filters is not well known. In fact, in the \maximai\ case we
have applied marginalization to deal with the lowest, $\simlt 0.1$~Hz, and the
highest frequencies, $\simgt 30$~Hz, for the reasons just mentioned. We
have found no strong dependence of our results on the specific choice
of bounds, obtaining nearly identical results if these values are
set to be $0.2$~Hz and $20$~Hz respectively.

\subsection{Singularities and pixel templates.}

\label{singularities:para}

Having accounted for these time stream `templates', $\bm{\tau_t}$, we
can produce the map and the {\em inverse} pixel-domain noise matrix. As
mentioned above, we may not be able to obtain the noise matrix itself,
due to zero eigenvalues in the inverse, corresponding to `infinite
noise' in some modes. 

However, just as the mapmaking procedure {\em per se} only requires
$\bm{N'_t}^{-1}$, subsequent manipulations of the map often can be cast in
terms of matrix inverses. (In another language, infinite noise
corresponds to zero weight.) For example, the likelihood function if we
consider a Gaussian-distributed signal is
\begin{equation}
  \label{eq:signallike}
  {\cal L}(C_\ell) = \frac{1}{\left|2\pi \bm{M_p}\right|^{1/2}} 
  \exp\left[ -\frac12 \bm{m_p}^T \bm{M_p}^{-1} \bm{m_p} \right],
\end{equation}
where $\bm{m_p}$ is the pixelized map as before, and $\bm{M_p}$ is 
the variance corresponding to the uncorrelated sum of the CMB
signal, $\bm{S_p}\l(C_\ell\r)$,
and the pixel-domain noise correlation matrix, \ie, 
$\bm{M_p}\equiv\bm{S_p}\l(C_\ell\r)+\bm{N_p}.$
If, due to time domain marginalization, $\bm{N_p}^{-1}$ has zero eigenvalue corresponding
to a pixel-domain template (cf.,~\eqn~\ref{mapsingmode:eq}), 
$\bm{v_p} = (\bm{A}^T\bm{N_t}^{-1}\bm{A})^{-1}\bm{A}^T \bm{N_t}^{-1} \bm{\tau_t},$ 
then we compute $\bm{N_p}$ as 
\begin{equation}
\bm{N_p}=\l(\bm{N_p}^{-1}+\epsilon\, \bm{v_p}\otimes \bm{v_p}^T\r)^{-1}-\epsilon^{-1}\bm{v_p}\otimes \bm{v_p}^T,
\label{singularinversion:eq}
\end{equation}
where $\epsilon$ is a small positive number and the inversion on the rhs is to be performed directly.
This procedure replaces an infinite eigenvalue of the $\bm{N_p}$ 
(corresponding to the eigenmode $\bm{v_p}$) with zero, leaving all other eigenvalues unaffected.
The total inverse correlation matrix is now to be understood as,
\begin{eqnarray}
  \label{eq:n+scorr}
  \bm{M_p}^{-1}&=&\lim_{\sigma_v^2  \rightarrow +\infty}\l(\bm{S_p}\l(C_\ell\r)+\bm{N_p}+\sigma_v^2\, \bm{v_p}\otimes \bm{v_p}^T\r)^{-1}\nonumber\\
  &=&\l(\bm{S_p}(C_\ell) + \bm{N_p}\r)^{-1}
  -\frac{\bm{\hat v_p}\otimes \bm{\hat v_p}^T} {\bm{v_p}^T \bm{\hat v_p}}, 
\label{fullsingularinversion:eq}
\end{eqnarray}
where 
$\bm{\hat v_p}\equiv \l(\bm{S_p}\l(C_\ell\r)+\bm{N_p}\r)^{-1}\bm{v_p}$ and 
the last term is a usual Sherman-Morrison-Woodbury term.
This expression assumes that $\bm{S_p}\l(C_\ell\r)+\bm{N_p}$ is invertible. However, 
any zero eigenvalue modes, additional to $\bm{v_p}$, can be treated in the same way
as $\bm{v_p}$, once a singular mode has been determined.

Even if we have not explicitly accounted for effects that may leave us
with a singular $\bm{N_p}$, we can use a similar technique, if we know
the pixel-domain pattern of the responsible modes. In this case, in
analogy with the time stream case (\eqn~\ref{tstream2:eq}), we can write the map as
\begin{equation}
  \label{eq:map+modes}
  \bm{m_p} = \bm{s_p} + a \bm{v_p} + \bm{n_p} = \bm{s_p} + \bm{n'_p}\; .
\end{equation}
Here, $\bm{s_p}$ is the signal, $\bm{n_p}$ is the noise, and $\bm{v_p}$
describes the shape of the unknown mode, with unknown amplitude $a$,
over which we will marginalize. Once again we can make an use of \eqn~(\ref{fullsingularinversion:eq}), 
with $\bm{N_p}$ this time computed in the standard way.

As an example, the total offset of the map is spurious and undetermined
(and it is often numerically convenient to set it to zero). That is, the
detectors are only sensitive to temperature variations, rather than
absolute temperatures. In the case of a lack of correlation between the
noise and the underlying map (as we have assumed all along), the inverse of the noise correlation
matrix computed for such a map should be singular by construction.  The
eigenvector corresponding to the zero eigenvalue is just a constant
function of a pixel number, \ie, $\bm{v_p}^T\equiv\l[1,...,1\r]^T$.
Knowing that, it is straightforward to perform the `inversion' as
in \eqn~(\ref{singularinversion:eq}).
In this specific case such
information can be included while computing the power spectra using
the MADCAP package~\cite{Borrill1999mad}.
An analogous problem has been addressed by the $COBE$-DMR team~\cite{Wright1996b}.
This equation can be straightforwardly generalized for cases with more
complicated  eigenvectors as discussed below.

\eqn~(\ref{singularinversion:eq}) simply sets to zero an infinite eigenvalue of an
inverse matrix. That formal procedure does not `solve' the problem of
singularity; rather it gives a compact expression for the noise
correlation matrix, which together with the knowledge of the singular
modes provides complete information needed for statistically sound
exploitation of the map. 

If such a singularity is identified, it usually can be dealt with
efficiently, producing a statistically valid result. Therefore,
determination of singular modes present in the map has to be a part of a robust map-making procedure.
Numerical inaccuracies often obscure singularities, 
making them difficult to find. The presence of such an undiscovered
singular mode does not
necessarily invalidate the outcome. In fact, in some
applications, the final result can be still correct, while such a
computation may accidentally
duplicate the (approximate) numerical marginalization technique discussed by
Bond, Jaffe \&\ Knox~\cite{BJK1998} and, \eg, implemented in
MADCAP. However, it is still advisable to first determine the singular modes
prior to applying such a method.

We also note in passing that this same formalism can be used to
`marginalize over' other sorts of template amplitudes at the map
stage, rather than the time stream templates considered earlier. An
important example of this are templates corresponding to known sources
of foreground emission, such as galactic dust, whose spatial morphology
is well-known from studies at other wavelengths, and are sub-dominant but
potentially important contaminants at CMB wavelengths~\cite{BJK1998}.

For other easily identifiable singularities see the next Section.

\subsection{Combining maps of time-stream segments}

\label{combining_segments:para}
\begin{figure*}[t]
\vskip -4truecm
\leavevmode\epsfxsize=18.cm\epsfbox{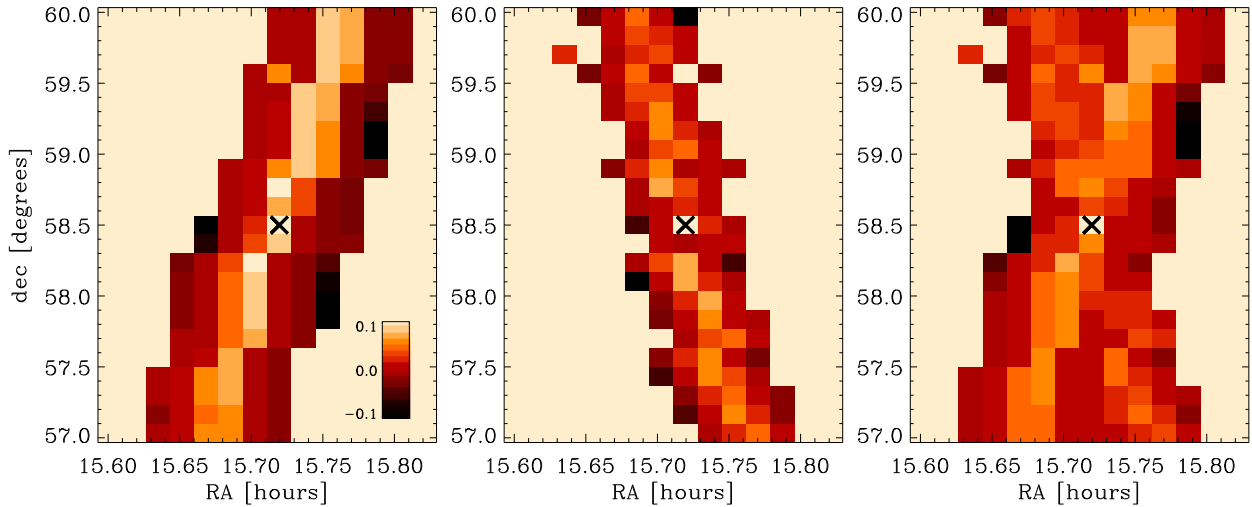}
\vskip -2truecm
\caption{Pixel domain correlations of the noise projected on the
sky. 
All three panels show a level of the correlations relative to the RMS
value of the noise for the same pixel, which is marked with an `x'. 
Color coding is shown in left panel. This pixel has been observed twice during the
\maximai\ flight. The noise correlations for the first observation are
shown in the left panel, and these for the second one
in the middle panel. Right panel shows the final co-added noise
correlations (see \eqn~\ref{matrixadding}). Clearly, due to the \maximai\
scanning strategy and the presence of the noise correlations in the time
domain, the noise correlation pattern in pixel domain is highly anisotropic and strongly
correlated as a result of any single observation of a pixel. The combined noise, for all pixels which were
observed twice, is however significantly less correlated and more isotropic.
\label{combnoise}
}
\end{figure*}
The map-making formalism as presented so far can be applied to a number of statistically
independent segments simultaneously. However, it may be 
advantageous to analyze each of these separately. In this case one needs to
combine the separate segments together at the end. For Gaussian noise
this would be quite straightforward, were it not for arbitrariness in
the offset of each segment. The latter
introduces possible relative offset shifts between the segments. This
can be resolved for  partially overlapping segments
if we require them to display, within the noise uncertainty, the
same underlying pattern in the common region. This introduces an
extra complication to the well-known formula (\eg,~\cite{daCosta1999}) for the
optimal co-addition of two maps. 
The maximum-likelihood problem can be solved in a standard manner or
using an approach analogous to that of Sect.~\ref{alternative:para}.
In consequence, on defining, for each time-stream segment,
$\bm{u_{p\l(I\r)}}$ as a pixel-domain vector of ones and
$\bm{\hat{u}_{p\l(I\r)}}\equiv \bm{N_{p\l(I\r)}}^{-1}\bm{u_{p\l(I\r)}}$, 
and introducing a corrected inverse noise
correlation matrix, $\bm{N'_{p\l(I\r)}}^{-1}$, such as (\cf, \eqns~\ref{margnoise}, \ref{fullsingularinversion:eq}),
\begin{equation}
\bm{N'_{p\l(I\r)}}^{-1}\equiv
\bm{N_{p\l(I\r)}}^{-1}-\frac{\bm{\hat{u}_{p\l(I\r)}}\otimes\bm{\hat{u}_{p\l(I\r)}}^T}
{\bm{u_{p\l(I\r)}}^T\bm{\hat{u}_{p\l(I\r)}}},
\end{equation}
we can express a final full map and a corresponding noise correlation
matrix in a familiar manner,
\begin{eqnarray}
\bm{N_p^{tot}}&=&\l\{\sum_I\bm{N'_{p\l(I\r)}}^{-1}\r\}^{-1},\label{matrixadding}\\
\bm{m_p^{tot}}&=&\bm{N_p}^{tot}\sum_I \bm{N'_{p\l(I\r)}}^{-1} \bm{m}_{\l(I\r)}.\label{segmentsadding}
\end{eqnarray}
Here  the sum is over maps of all segments to be combined.
As discussed above 
the undetermined absolute offsets of each of the segments separately is
reflected in the singularity of each of the redefined inverse noise
correlation matrices, ${\bm{N'_{p\l(I\r)}}}^{-1}$.
The final inversion in \eqn~(\ref{matrixadding}) again has to be
understood as in \eqn~(\ref{singularinversion:eq}) and the singularity needs
to be accounted for in subsequent stages of the data analysis, as, for
instance, shown in \eqn~(\ref{fullsingularinversion:eq}).

A \maximai\ based example of an application of this procedure is shown in
Fig.~\ref{combnoise}. As in the case of a single continuous time stream
segment~\cite{Tegmark1997design}, re-observing the same patch of the sky
along the different scanning direction not only suppresses the level of
the noise per pixel but also weakens and isotropizes the correlation
pattern in pixel domain.

Note that the expression in the curly brackets of \eqn~(\ref{matrixadding}) becomes
singular also, if not every segment map is connected (directly or
through the number of intermediaries) with the others.
If no such link exists the relative offset of such a segment map
with respect to the rest of the map remains unknown giving rise to
a singular mode in addition to the one related to the absolute
offset. However, \eqns~(\ref{matrixadding}) \&\ (\ref{segmentsadding}) can still be applied if
the inversion of the singular matrix is interpreted as in
\eqn~(\ref{singularinversion:eq}). The singular eigenvector $\bm{v_p}$
has to now be appropriately replaced.

For instance, in a case of a single disjoint segment the two singular modes
emerging as a result of the unknown total offset of the full map and a
relative offset of the disjoint part can be chosen as having ones for
every pixel belonging to one disjoint part and zeros elsewhere (or
vice verse).
With the inversion now described by the straightforward generalization
of \eqn~(\ref{singularinversion:eq}) for the case of multiple singular eigenvectors,
the final product of the operation given by 
\eqn~(\ref{segmentsadding}) can be a map composed of many disconnected
regions with the uncertainty due to our ignorance of their relative offsets incorporated
into the total noise correlation matrix, $\bm{N_p^{tot}}$, as given by \eqn~(\ref{matrixadding}).

Numerically one may encounter nearly singular cases whenever the
overlapping region between two segments is too limited or the noise per
pixel in the overlapping area too high to provide any useful constraint on the free offset.
A practical and safe way of dealing with such a 
problem is to reject a (small by assumption) number of common pixels to 
make a given part genuinely disconnected and to account in a mathematically
strict manner for the arising singularity of the noise correlation matrix.

A power spectrum of such an unconnected map can be subsequently computed
as explained in Sect.~\ref{singularities:para}.
However, that requires more involved algebra than
that currently implemented in the MADCAP version of the quadratic estimator. For that reason we have rejected 
all the disconnected segments
from the \maximai\ maps while computing their power spectra. In that way the number 
of segments used for the final analysis decreased to 14 per detector~\cite{Hanany2000,Lee2001}.

\subsection{Combining maps of different photometers.}

The additional difficulty here in comparison with the previous Section 
is introduced by the possible relative calibration difference between maps produced for 
different detectors. 
Calibration uncertainty is generically difficult to be included into a
maximum likelihood framework due to its multiplicative character.
For the \maximai\ data set we have found that the relative calibration 
between different photometers' maps are with a high precision correct if the mean dipole based 
calibration is adopted for each of the maps~\cite{Stompor2001b}. Therefore rather than seeking a general solution
to the problem,  we combined maps as given  by the \eqn~(\ref{segmentsadding}) and used the largest single
detector error for the calibration uncertainty of the combined map. 

\subsection{Low-$\ell$ aliasing.}

\label{cleaning:para}

A potential bias of the final anisotropy power spectrum 
(but also for other statistics, see, \eg, \cite{Cayon2001}) resulting 
from
this kind of the data analysis is due to an incomplete (and modest in a
case of all balloon-borne experiments) sky coverage. That induces the
correlation between otherwise uncorrelated (for a statistically
isotropic sky) $\ell$ modes.  As a consequence, power contained in the
low-$\ell$ modes beyond the detection capability can leak to the higher
$\ell$ modes which are to be estimated. Because the amplitude of the
anisotropy power spectrum usually decays with increasing $\ell$ -- and
is many orders of magnitude higher in monopole and dipole than in any
other mode -- there is a potential for biasing of the low-$\ell$ end of
the estimated power spectrum. 

One possible approach is to consider the unwanted modes as pixel-space
templates as discussed above (Sect.~\ref{singularities:para}): 
$\bm{v_p}\l(i_p\r)=Y_{\ell m}(i_p)$ for all the
$(\ell,m)$ modes to be marginalized over. ($i_p$ is here a pixel number.) We then explicitly marginalize over
them, while estimating, \eg, the anisotropy power spectrum
(\eqn~\ref{eq:n+scorr}).
This is closely analogous to the time stream frequency marginalization of Sect.~\ref{marginalization:para}.

Another, approximate, way to deal with the
problem also can be applied directly on the power spectrum estimation stage
is implemented in the MADCAP package~\cite{Borrill1999mad,Borrill2000}
as described at the end of this Section.

An alternative exact solution is based on  G{\'o}rski's
idea~\cite{Gorski1994}. In this approach the unwanted low-$\ell$ modes are removed from
the map prior to further statistical analysis and the corresponding
noise correlation matrix in pixel domain is appropriately corrected to
account for the additional uncertainty. Unlike the other just-mentioned
options this method produces a `cleaned' version of the
sky map to be used henceforth.

Let us start by defining a scalar product of two functions $\bm{f}$ and $\bm{g}$ defined
at each pixel $i_p$ of our map $\bm{m_p}$ as (hereafter $\star$ stands
for a complex conjugate),
\begin{equation}
\l( \bm{f} | \bm{g}\r) \equiv \frac{1}{n_{pix}}\sum_{i_p} \bm{f}^\star\l(i_p\r)\bm{g}\l(i_p\r).
\end{equation}
Also we denote by $\l\{\bm{y}\r\}$ a subset of the $\l(\ell_0+1\r)^2$ spherical harmonics of the order not
higher than $\ell_0$ which are to be removed form the map.
In general $\l\{\bm{y}\r\}$ is neither a
linearly independent nor a complete basis on the map $\bm{m_p}$. However,
we can construct a set of orthonormal functions $\l\{\bm{\psi}\r\}$
spanning the space of the spherical harmonics included in $\l\{\bm{y}\r\}$.
The construction can be performed by using Singular Value or
Cholesky decomposition of the Kowalewski-Gram
determinant of $\l\{\bm{y}\r\}$ functions (as, \eg,~\cite{Gorski1994}) and
rejecting all the singular modes.
The resulting set of functions though orthonormal is clearly not complete. To achieve completeness
we can supplement it with the functions from the another orthonormal (and complete) set 
of functions defined on the map and retain from the latter -- through a standard Gram-Schmidt 
orthonormalization procedure -- only those functions (or their linear combination) which are orthogonal 
to all the $\l\{\bm{\psi}\r\}$ functions. 
Practically, that part of the procedure can be encoded using Singular
Value Decomposition.
A convenient choice of the extra complete functional basis is just a
pixel basis, $\l\{\bm{p}\r\}$,
\begin{equation}
\bm{p_{\l(i\r)}}\l(j_p\r)\equiv\delta^K\l(i,j_p\r), \mbox{where  } i,j_p=1,...,n_{pix}.
\end{equation}
On the successful completion of the entire procedure we end up with the set of the $n_{pix}$ orthonormal functions.
By construction, all the $\l\{\bm{\psi}\r\}$ functions are included in the final basis, and they span all the spherical
harmonics of the order $\le \ell_0$ on the map $\bm{m_p}$.
Hence all the remaining functions
of the final basis (denoted hereafter $\l\{\bm{\xi}\r\}$), which are
orthogonal to the functions $\bm{\psi}$ by construction, are also orthogonal to
the all spherical harmonics of that order. 

Algebraically the described procedure consists of a pair
of linear transformations, which in a pixel representation can be written down as,
\begin{eqnarray}
\bm{\psi}&\equiv&  \bm{K} \bm{y},\\
\bm{\xi} &\equiv & \bm{L}\bm{p}\label{orthobasis:eq}
\end{eqnarray}
Here $\bm{K}$ changes the basis from that of spherical harmonics,
$\l\{\bm{y}\r\}$, to $\l\{\bm{\psi}\r\}$.
$\bm{L}$ transforms the pixel
basis, $\l\{\bm{p}\r\}$, to the orthogonal basis made of basis vectors
either parallel or perpendicular to $\l\{\bm{\psi}\r\}$ and retains
only the latter subset, $\l\{\bm{\xi}\r\}$. These functions
complement $\l\{\bm{\psi}\r\}$ and both sets together form an orthonormal and complete
basis $\l\{\bm{\psi},\bm{\xi}\r\}$ on the map, such as 
\begin{equation}
\l(\bm{y}|\bm{\xi}\r)=0.
\end{equation}
The map purged of all the contribution from the low ($\le \ell_0$) order spherical
harmonics is then given as,
\begin{equation}
\bm{m'_p}=\sum_j \l(\bm{m_p}|\bm{\xi_{\l(j\r)}}\r)\bm{\xi_{\l(j\r)}},
\end{equation}
and is therefore uniquely represented by a vector $\bm{m_\xi}$ defined
as,
\begin{equation}
\bm{m_\xi}\l(i\r)\equiv\l(\bm{m_p}|\bm{\xi_{\l(i\r)}}\r).
\end{equation}
The total (signal plus noise) correlation matrix for $\bm{m_\xi}$ is then
given as,
\begin{equation}
\langle \bm{m_\xi} \otimes\bm{m_\xi}^T\rangle=
\l[\bm{L}\l(\bm{S_p}+\bm{N_p}\r)\bm{L}^T\r],\label{corrected_corr}
\end{equation}
where $\langle ...\rangle$ denotes an ensemble average.
$\bm{S_p}$ and $\bm{N_p}$ are signal and noise correlation matrices computed
for the complete map, $\bm{m_p}$, prior to any mode removal,
\ie, $\bm{S_p}+\bm{N_p}\equiv \langle \bm{m_p} \otimes\bm{m_p}^T\rangle$.
If the sky signal contains only CMB anisotropy then the signal term, $\bm{S_p}$, is given by,
\begin{equation}
\bm{S_p}=\frac{1}{4\pi} \sum_{\ell} \bm{P_\ell} \l(2\ell+1\r) C_{\ell},
\label{fullcorr:eq}
\end{equation}
where $C_\ell$ is the power spectrum of the CMB fluctuations, 
and $\bm{P_\ell}$ denote matrices of the Legendre polynomials computed
for all the pairs of the pixels in the given map, \ie,
\begin{equation}
\bm{P_\ell}\l(i_p,j_p\r)\equiv P_\ell\l[\cos\l(\bm{\gamma}\l(i_p\r)\cdot\bm{\gamma}\l(j_p\r)\r)\r].
\end{equation}
Here $\bm{\gamma}\l(i_p\r)$ is a unit vector pointing at the center of the
pixel $i_p$ and $P_\ell$ a standard Legendre polynomial of an order $\ell$.

The correlation matrix of the map without low-$\ell$ modes, as defined in 
\eqn~(\ref{corrected_corr}), 
can also be rewritten as a sum of a noise-like, $\bm{N_\xi}$, and
signal-like, $\bm{S_\xi}$, term,
\begin{equation}
\bm{L}\l[\bm{S_p}+\bm{N_p}\r] \bm{L}^T=\bm{S_\xi}+\bm{N_\xi}
\label{cleanedcorr:eq}
\end{equation}
Furthermore, on introducing the redefined Legendre polynomials matrices 
$\bm{P'_\ell}\equiv\bm{L}\bm{P_\ell} \bm{L}^T$ and using \eqn~(\ref{fullcorr:eq}), we can rewrite the
signal part of \eqn~(\ref{cleanedcorr:eq}) as,
\begin{eqnarray}
\bm{S_\xi}&=&\frac{1}{4\pi} \sum_{\ell} \bm{L} \bm{P_\ell} \bm{L}^T
\l(2\ell+1\r) C_{\ell}\label{cleanedcorr2:eq}\\
&=&\frac{1}{4\pi} \sum_{\ell} \bm{P'_\ell} \l(2\ell+1\r) C_{\ell}.\nonumber
\end{eqnarray}
The last equation shows that the signal correlation matrix of the 'cleaned' map is related to 
the angular power spectrum of 
the CMB anisotropies, $C_\ell$, in a way which is formally identical to that for the complete map,
\cf, \eqn~(\ref{fullcorr:eq}).
Therefore we can use usual quadratic estimator algorithms and, in particular, 
the MADCAP package~\cite{Borrill1999mad,Borrill2000} to estimate the CMB power spectrum
having at disposition only the map (and the corresponding noise correlation matrix, \eqn~\ref{cleanedcorr:eq}) with low-$\ell$ modes 
removed.
To use that package in this case one just needs to replace
matrices $\bm{P_\ell}$ (which constitute an intermediate output of
MADCAP) with matrices $\bm{P'_\ell}$ and use the cleaned map in the $\bm{\xi}$ representation, $\bm{m_\xi}$, and
the %corrected 
noise correlation matrix, $\bm{N_\xi}$, as an input, instead
of usual $\bm{m_p}$ and $\bm{N_p}$.

Though the matrices $\bm{K}$ and $\bm{L}$ can be chosen to be sparse
(nearly triangular) the entire procedure is quite involved and the computational cost scales as 
\order{n_{pix}^3}. The same scaling applies to the computations of 
products of $\bm{L}$ and $\bm{P_\ell}$ matrices.
However, while only the low
$\ell$ part of the angular power spectrum is likely to be affected by
that kind of aliasing, the results of such a test applied to a map with
relatively large pixels can provide a useful estimate of the size of
the expected effect. In the \maximai\ case
we applied this test to the map with $10'$ pixels (\ie, less than $4000$
pixels in total) and compared it with the simple approximate template
procedure of \cite{BJK1998} as implemented in the MADCAP package of weighing out 
the monopole and dipole from
the map, and introducing an extra low-$\ell$ bin to the recovered anisotropy power spectrum
which is {\em a posteriori} rejected. In the \maximai\ case
this bin extended from $\ell=2$ up to $\ell=34$.
Its rejection corresponds to the marginalization over
that bin power and would be formally strict only if the likelihood for the power spectrum bin amplitudes
were precisely Gaussian. Though such an approach is less general, 
because of these approximations and because it explicitly assumes the isotropy
of the sky signal in the rejected modes,
we found no difference between the results of both methods.
Therefore, we conclude that the \maximai\ final anisotropy power spectrum,
in the published to date range of multipoles, $35\simlt \ell \simlt 1235$,
is not 
affected by any appreciable contribution due to aliasing of the
low-$\ell$ mode power.

\section{Iterative noise estimation.}
\label{iteration:para}

\begin{figure*}[t]
\vskip -4truecm
\leavevmode\epsfxsize=18.cm\epsfbox{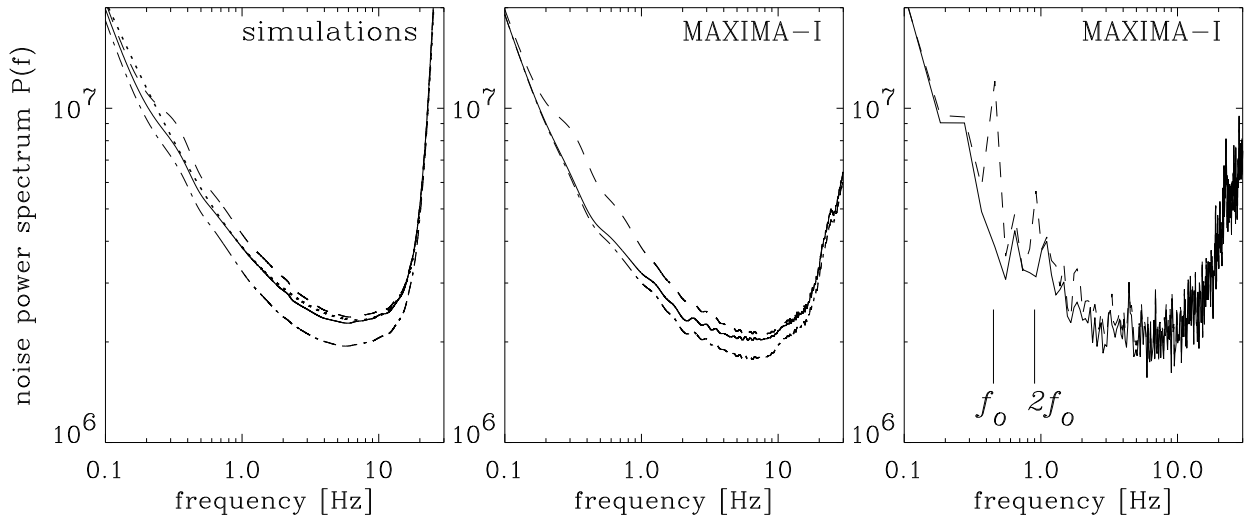}
\vskip -2truecm
\caption{Estimated noise power spectra for simulated and real data
using the iterative approach of Ferreira \&\ Jaffe~\cite{Ferreira2000}.
In left panel, dashed line shows a noise power spectrum computed
as in Sect.~\ref{noisestim:para} without iterative corrections. The dotted line
depicts the noise spectrum used for a simulation, that almost perfectly
overlaps with a solid line corresponding to a noise power spectrum calculated
using 4 iterative steps and a map with 8' pixels. The dash-dotted line
shows a result after 4 iteration but using a map with 3' pixels. That
demonstrates the bias as described in the text.
Middle panel shows noise power spectra for a single segment of the
actual \maximai\ data. The power spectrum has been recovered
using various approaches as in the left panel: the solid line shows the
spectrum computed with 4 iterative steps and 8' pixels, dashed line shows the
uniterated estimate, and dash-dotted line shows a result of 4 iterative
steps with 3' pixels.
The right panel shows noise spectra
obtained just as a result of averaging on the second step of the noise
estimation procedure, prior to any smoothing (see
Sect.~\ref{noisestim:para} and the
right most panel of Fig.~\ref{noisestim:figure}). 
Dashed line corresponds to a spectrum obtained with no iteration, displaying 
characteristic two small spikes at $\sim 0.45$Hz and $0.9$Hz corresponding to a primary mirror chop
frequency and its first harmonic. The overplotted solid line is a
spectrum after 4 iteration using 8' pixels. Clearly in the latter
spectrum the spikes are corrected as a result of iterations. 
}
\label{noisiter:figure}
\end{figure*}

To date \maxima\ is one of the most sensitive experiments in terms
of the noise level per measurement achieving the level of 
$\sim 1500 \mu$K, \ie, $100\mu \mbox{K} \sqrt{\mbox{s}}$,  for some of the detectors.
In spite of that the expected CMB and foreground signal for the
observed patch of sky is still expected to be a sub-dominant part
of a single measurement ($\ll 10\%$ of the total power). 
The sky-related contribution to the power confined within some of the frequency bands
can be, however, much higher ($\simlt 10\%$).
Similarly, the non-stationary effects,
though they may appear to be quite small,
they can be limited to a number of narrow frequency bands,
dominating the power in there.
For instance, in the \maximai\ case the primary mirror synchronous
signal with its amplitude of $200\mu$K dominates occasionally the power
in the narrow frequency bands centered on the fundamental mode of primary mirror chop and its
few lowest harmonics (see Fig.~\ref{noisiter:figure}).

The important assumption behind the noise power spectrum estimation
as presented in Sect.~\ref{noisestim:para}, \ie, that noise dominates the time stream
measurements, though not clearly breached needs to be, therefore, tested.

To account for that effect we follow the iterative approach of~\cite{Ferreira2000}.
It attempts to recover the noise component
of the entire time stream, which is subsequently 
used in the noise estimation procedure (see Sect.~\ref{noisestim:para}).
The starting point of the iterative procedure is an approximation that
$\bm{n_t}^{\l(0\r)}\simeq\bm{d_t}$.
Given this, it proceeds to the noise estimation and then to the map making. The resulting (zeroth order) map 
is used as an estimate of the signal in the time stream and is subtracted from the time stream on the
next iterative step~\cite{Ferreira2000,Prunet2000}.
If we denote the noise contribution to the time stream on the $i$th step as
$\bm{n_t}^{\l(i\r)}$ then,
\begin{equation}
\label{noise_iteration}
\bm{n_t}^{\l(i\r)}=\bm{d}-\bm{A}\bm{m}^{\l(i-1\r)}-\bm{B}\bm{x}^{\l(i-1\r)}=
{\cal{A}}^{\l(i-1\r)}\gm^{\l(i-1\r)},
\end{equation}
where $\bm{m}^{\l(i-1\r)}, \bm{x}^{\l(i-1\r)}$ and
$\gm^{\l(i-1\r)}$ are a map and  a primary mirror signal and a generalized map respectively as estimated on 
the previous step. As shown by ~\cite{Ferreira2000} 
the differences between the maps and noise correlation matrices estimated on the subsequent steps of the 
iteration decrease very quickly and the required precision is achieved rather rapidly.
Usually we have found that at most four iterative steps were needed to reach the accuracy of few percent.
The iterative approach can be 
significantly sped up if no explicit inversion of the pixel-pixel noise correlation matrix is performed 
on each iterative step but the map is calculated using an iterative linear system solver
~\cite{Prunet2000,Prunet2001,Dore2001a}. 

It is important to notice that the method is only asymptotically,
-- \ie, in the limit of the large number of effective degrees of
freedom -- unbiased as guaranteed by its maximum likelihood origin. 
That limit is achieved, for instance, when a number of time samples increases, but
a number of pixels is fixed.  

The ensuing bias can be estimated as follows. Let us assume that we
know the noise correlation function, $\bm{N_t}$. Though the noise
estimation goal seems to have been achieved, the iteration, as given by
\eqn~({\ref{noise_iteration}}), can go on.
The time stream estimate on the next step will contain noise only
but composed of two components: a true time stream noise $\bm{n_t}=\bm{n_t}^{\l(i\r)}$ and
a pixel domain noise, $\bm{n_p}$, projected back to the time domain via the
pointing matrix (and therefore also non-stationary):
\begin{equation}
\bm{n_t}^{\l(i+1\r)}=\bm{n_t}^{\l(i\r)}-{\cal A} \bm{n_p}=\bm{n_t}-{\cal A} \bm{n_p}.
\label{FJbias}
\end{equation}
The correlations of the latter quantity are computable and given by
\begin{eqnarray}
\l\langle \bm{n_t}^{\l(i+1\r)}\l(i_t\r)\bm{n_t}^{\l(i+1\r)}\l(j_t\r)\r\rangle=
\bm{N_t}\l(i_t,j_t\r)&&\label{corrbias}\\
-\sum_{i_p,j_p}{\cal A}\l(i_t,i_p\r) {\cal N}_{p}\l(i_p,j_p\r) {\cal A}^T\l(j_t,j_p\r).&&\nonumber
 \end{eqnarray}
Here the summation is over all pairs of pixels 
and ${\cal N}_{p}$
stands for the noise correlation matrix in the (generalized) pixel domain
corresponding to the true noise correlation in time domain $\bm{N_t}$.
$\langle ...\rangle$ denotes an average over the statistical ensemble
of the noise realizations.

Interestingly, due to existing correlations between noise in the time
and pixel domain `adding' the extra noise to the time stream as
described by \eqn~(\ref{FJbias}) results in the underestimation of
the actual noise power in the time domain. 
Though the above formula does not really help to unbias
the procedure it gives a useful criterion of its applicability.
For practical use it is useful to consider the overall
power suppression due to the bias. Taking the trace of \eqn~(\ref{corrbias})
we get,
\begin{eqnarray}
 \sum_{i_t}\l\langle \bm{n_t}^{\l(i+1\r)}\l(i_t\r) \bm{n_t}^{\l(i+1\r)}\l(i_t\r)\r\rangle
&=&\label{powerbias} \\  \sum_{i_t} \bm{N_t}\l(i_t,i_t\r)
&-& \sum_{i_p}n_{t}\l(i_p\r) {\cal N}_{p}\l(i_p,i_p\r).\nonumber
\end{eqnarray}

Two limiting cases are evident. If each of the pixels is observed
once then the noise iteration has no meaning, as we have no means to
distinguish between the noise and the sky signal, and its result is fully
biased, \ie, the left hand side of the above equation is zero. If there is
no correlation in the time domain then pixel noise is just
proportional to the number of times a given pixel has been
observed. Inserting that into \eqn~(\ref{powerbias}) renders,
\begin{equation}
\frac{1}{n_s} \sum_{i_t}\l\langle \bm{n_t}^{\l(i+1\r)}\l(i_t\r) \bm{n_t}^{\l(i+1\r)}\l(i_t\r)\r\rangle
=\sigma_{t}^2\frac{\l(n_s-n_{pix}\r)}{n_s},
\end{equation}
where $\sigma_{t}^2$ is a diagonal element of $\bm{N_t}$.
This formula, which in the white noise case can be also derived
directly from the maximum likelihood considerations,
quantifies the introduced fractional bias as equal to $n_{pix}/n_s$.
If the number of pixels is fixed then in the limit of the increasing
number of measurement the bias disappears as expected. 

If the correlations in the time streams are not negligible then a
similar expression can be concocted with the number of pixels
replaced by an `effective' number of the parameters $n_{eff}$, which
is to be determined case by case,
\begin{equation}
\frac{1}{n_s} \sum_{i_t}\l\langle \bm{n_t}^{\l(i+1\r)}\l(i_t\r) \bm{n_t}^{\l(i+1\r)}\l(i_t\r)\r\rangle
\simeq \sigma_{t}^2 \frac{\l(n_s-n_{eff}\r)}{n_s},
\end{equation}
The intuitive meaning of $n_{eff}$ is clear from the expression
\begin{equation}
n_{eff}\equiv \sum_{i_p} n_s\l(i_p\r) \frac{{\cal N}_{p}\l(i_p,i_p\r)}{\sigma_{t}^2}
\label{biasestim}
\end{equation}
In the \maximai\ case $n_{eff}$ is not smaller than $n_{pix}$, and hence
correlations tend to increase the bias of the noise iteration procedure.

It is important to notice that due to our assumption of stationarity,
the bias depends only on how on average the measurements are divided
between pixels and is therefore robust to the presence of the
poorly sampled pixels.

From \eqn~(\ref{powerbias}) it is clear that the bias is a result of
the noise presence in the pixel domain. It is therefore advantageous to
use all available information to minimize the noise of the estimated
map, including all the time stream data during which 
a given patch of the sky has been observed.

In the \maximai\ case we have found (Fig.~\ref{noisiter:figure}) that
it is still advantageous to perform noise iterations for pixels as small
as 8 arcminutes. For smaller pixels, the resulting map is occasionally too noisy,
and the likely bias larger than an expected gain. Therefore, in those cases,
we either restrict iterative corrections to deal only with the
primary mirror synchronous signal or use noise power spectrum 
estimates obtained for the 8 arcminute pixels~\cite{Lee2001}.
The latter approach is helpful in correcting for the low-angular-scale
power, 
which should constitute the bulk of the sky signal
present in the time stream. However, one may worry that it also
introduces spurious correlations at the time lags corresponding to the
characteristic crossing time of the big pixel.
In practise, we have found that the results rendered by both these
approaches are in very good agreement.

Clearly, invoking some kind of a map-denoising method and/or applying
on this stage more aggressive filtering can be useful
to extend the applicability of iterative noise estimation.
We leave this issue for future research.

\section{Consistency tests.}

\label{constest:para}

\begin{figure*}[t]
\vskip 0truecm
\leavevmode\epsfxsize=16.cm\epsfbox{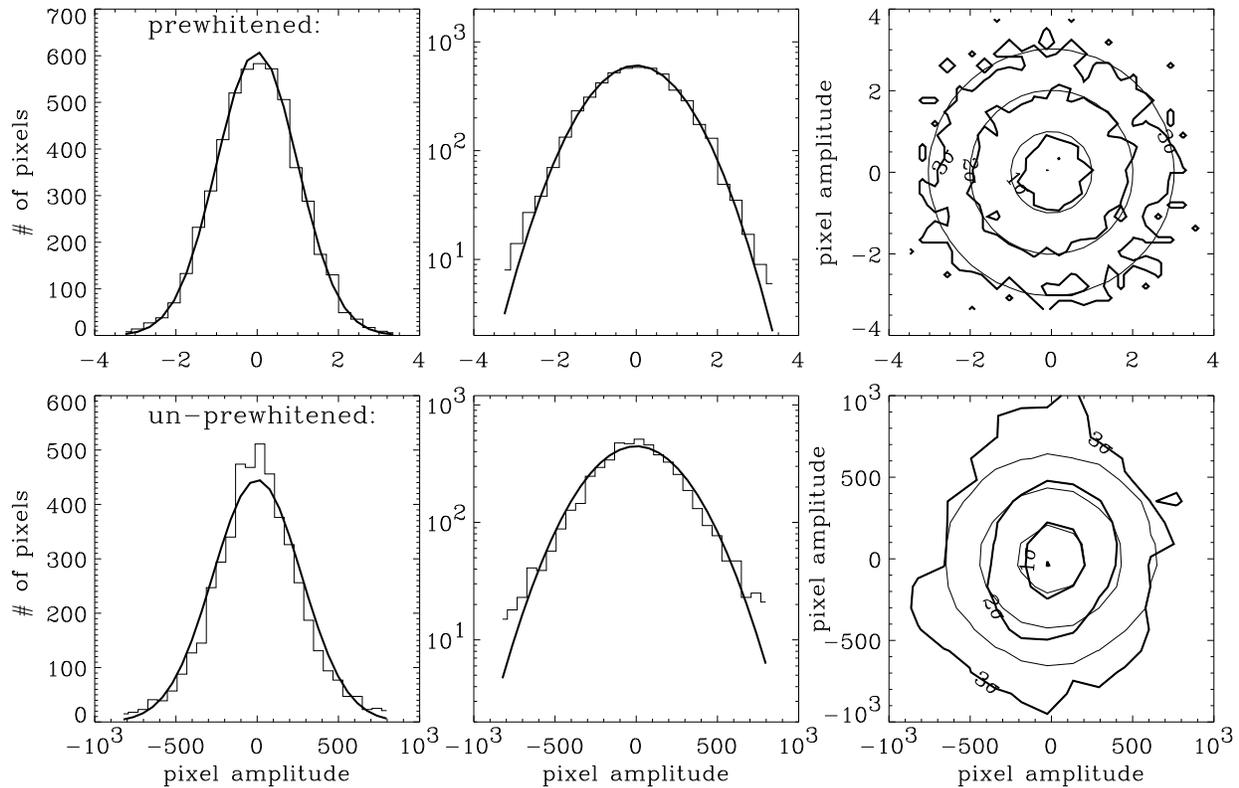}
\vskip -1truecm
\caption{
Probability distribution functions of the pixel amplitudes for the noise maps computed as a
difference of two maps made for two different photometers. The upper row
shows the results using prewhitened maps. Their corresponding
unprewhitened versions are displayed in the lower row. Histograms
show the results for the real data and smooth lines the Gaussian
curves with a dispersion equal to one (upper panels) or fitted to best
match the actual results.
The two right panels show contour plots of the histograms of joint two-dimensional
probability distributions of the pixel temperature for two `noise' maps.
These were computed as a difference of the sky maps recovered 
for two independent pairs of the photometers. The lower panel shows the results for
the actual maps with the correlated pixel noise, and upper panel for the
maps prewhitened prior to histogramming. Concentric ellipses marked
with thin solid lines show the Gaussian expectation. Those were
computed assuming no correlations between both maps and were
discretized the same way as data histograms and setting the dispersions
either to unity (upper panel) or to the best fit values (lower panel).
\label{ksall:figure}
}
\end{figure*}

A number of assumptions and approximations are involved
in the computations of a map and a corresponding noise correlation
matrix, so it is desirable to test the consistency of the
final map-making products. Clearly that
is difficult for the maps containing still-to-be-determined
cosmological contributions without any prior assumptions. However, it can be
done for the maps containing just noise. Those can be either the
projections on the sky  of the so called `dark' bolometers usually incorporated
in experiments
to track instrumental effects in the data~\cite{Lee1999}, or just differences
of the maps computed for single detectors.
If, as expected, the maps recovered for a single photometer contain only the sky
signal and the noise, the subtraction removes
the sky component leaving a map of the noise only.
Under the assumption of Gaussian time stream noise, the noise maps are
also Gaussian, and their correlations are given by the noise
correlation matrices produced in parallel by a map-making procedure (Sect.~\ref{mapmaking_main:para}).
 Any failure to meet such a requirement would suggest
either a failure of the basic assumptions or some other
problem with the data, the data analysis methods, or both.

The noise maps, due to correlations and inhomogeneity of the noise,
are described by the multi-dimensional probability distributions
with a number of dimensions equal to a number of pixels $n_{pix}$ of a
map under consideration.
Therefore, notwithstanding the large number of pixels, 
a simple histogram of the pixel
temperatures is likely to display significant deviations from the
1-dimensional Gaussian distribution even for the truely Gaussian case.
To alleviate this problem we first `prewhiten' the map performing a
linear transformation in order to decorrelate the measurements in the
different pixels of the map,
\begin{equation}
\bm{w_p}\equiv \bm{N_p}^{-\frac{1}{2}}\bm{m_p},
\end{equation}
here $\bm{N_p}^{-1/2}$ is a `square root' of the noise correlation
matrix as estimated from the map-making, satisfying a relation, 
\begin{equation}
\bm{N_p}^{-1}\equiv \bm{N_p}^{-{1/2}}\l[\bm{N_p}^{-{1/2}}\r]^T.
\end{equation}
In the following we take $\bm{N_p}^{-{1/2}}$ to be a Cholesky triangular matrix
(\eg,~\cite{Golub1983}).
If the noise correlations are estimated correctly (or at least consistently), then the components
of the vector $\bm{w_p}$ (a `prewhitened map') are uncorrelated and
their variances are equal to unity. Thus the multi-dimensional problem
reduces to the well-defined one-dimensional test.
Moreover, if the noise in the pixel domain is Gaussian
then each of the components of $\bm{w_p}$ is randomly drawn from the Gaussian
1-dimensional distribution with the unit variance.
That is the hypothesis which we test. We apply a one dimensional
Kolmogorov-Smirnov test. Its results give an estimate of how often the
one point distribution function such as the one actually measured 
can be obtained from the Gaussian distribution
with a unit variance. For \maximai\ we have four single-detector maps,
giving us six independent difference maps. We test all of these noise maps, finding
that the KS significance is always higher than $\sim 10$\% 
and usually as high as $\sim 50-60$\%, confirming the very good
consistency of our map-making products. The sample of the results is also shown in 
Fig.~\ref{ksall:figure}. Clearly, a histogram of the unprewhitened map shows a significant
deviation from the Gaussian curve near the peak (lower, left panel of the figure) and in the tails
of the distribution. As anticipated, prewhitening largely resolves the discrepancies, allowing us to
recover nearly perfectly Gaussian curves.

Though such KS-like tests usually provide a weak diagnostic, they are powerful
consistency tests when passed. Sources of possible failure are abundant. Those can be either
problems of the data set like an extra photometer-dependent parasitic signal in the
time-stream leaving its imprint in the difference maps, or
non-Gaussianity of the time domain noise, or cross-correlation between
maps used for the creation of the difference maps. 
Moreover, the fact that one of our maps was actually
made of the data obtained by the photometer centered at a higher frequency
($\sim 240$ GHz) than the others frequency ($\sim150$ GHz) also
suggests the lack of a
substantial frequency-dependent sky (\eg, foreground) signal, as 
expected from the choice of the low contamination contrast patch
observed during the \maximai\ flight.
Alternatively, the
source of the problem may lay with the data analysis, such as
noise misestimation both in the time domain and pixel domain as a
result of the involved approximations. 
In fact we have found that the results of KS tests depend on the precise map-making procedure we choose to apply to
the real data. In particular, we derive somewhat different numbers if, \eg, no noise iteration has been 
performed, or no primary-mirror-synchronous signal has been removed. In both cases the differences are mainly
due to the differences in the long wavelength modes present in the
maps, which are the most susceptible to the details of the map-making algorithm.
Nevertheless, that shows that the KS test possesses 
sensitivity which makes it a useful tool in tracking realistic
problems in the maps and/or procedure.
The KS test can be also applied to the maps with the low-$\ell$
multipoles removed as described in the previous Section.
Hence the lack of an
indication of the problem with the KS test results is quite
encouraging and may serve as a fairly comprehensive validation of the
final results. 

In addition, for every pair of the noise maps we can consider a
two-dimensional probability distribution of two noise maps calculated
as differences of two actual single detector maps. Here we use a
difference map of the first and second detector, $\bm{m_{p\l(12\r)}}$,
and of the third and fourth, $\bm{m_{p\l(34\r)}}$ and consider the
probability distribution ${\cal P}\l(\bm{m_{p\l(12\r)}},\bm{m_{p\l(34\r)}}\r)$.

If we assume that the signal detected by different detectors is
uncorrelated, the correlation matrix for a pair $\l(\bm{m_{p\l(12\r)}},\bm{m_{p\l(34\r)}}\r)$
has a block diagonal structure with blocks given by the correlation matrices of each
of the difference map separately. Prehwitening in such a case is simply given by,
\begin{equation}
\l[
\begin{array}{c}
{\dsp \bm{w_{p\l(12\r)}}}\\
{\dsp \bm{w_{p\l(34\r)}}}
\end{array}
\r]\equiv \l[
\begin{array}{c c}
 {\dsp \bm{N_{p\l(12\r)}}^{-1/2},} & \bm{0}\\
 \bm{0}, &  {\dsp \bm{N_{p\l(34\r)}}^{-1/2}} 
\end{array}
\r] \l[
\begin{array}{c}
{\dsp \bm{m_{p\l(12\r)}}}\\
{\dsp \bm{m_{p\l(34\r)}}}
\end{array}\r].
\label{ks2dim}
\end{equation}
The two dimensional probability distribution ${\cal P}\l(\bm{w_{p\l(12\r)}},\bm{w_{p\l(34\r)}}\r)$
is then bound to have a unit variance and the Gaussian shape unless
the cross-correlation term is indeed present or any of the previously
mentioned reasons/assumptions is not fulfilled.
The obtained results (right hand panels of Fig.~\ref{ksall:figure}) seem to agree well with the expectations and therefore
confirming the consistency of our analysis.

\section{Summary.}
Recent CMB data sets have set new challenges for the data
analysis. Problems are related to a sheer
size of new data sets and also to a quality of the analysis tools,
which can make a full use of increasing power of data.

This paper describes how those challenges were met in the analysis of the \maximai\ data set.
The successful production of the final results, as published in the recent papers~\cite{Hanany2000,Lee2001},
required us to improve on the existing and develop and test new methods and tools. 
Though some of them had to be significantly
customized to be efficient, the others seem to be of a more general character and
applicability extending to data sets as big as that of the forthcoming satellite missions ~\cite{Map,Planck}.

We have focused here on time ordered data manipulation techniques
and map-making algorithms. We have presented a comprehensive, consistent
approach allowing us to recover a map of the sky and to estimate its
error matrix in realistic circumstances of an actual CMB experiment.
The highlights include: 
\begin{itemize}
\item{the time stream noise estimation procedure coupled together with the 
gap filling method through constrained noise realization;}
\item{an exact version of the map making code;}
\item{the statistically sound methods of dealing with the time stream filtering of 
contaminated frequencies and time-domain templates.} 
\end{itemize}
The entire suite of the methods  presented here
amounts to a self-consistent approach to building a final map out of
smaller parts in an efficient way. The construction can be halted when the largest map
lending itself to the statistical analysis (\eg, power spectrum estimation) has been built.
Dealing with only subsets of an entire data set seems to be the only
efficient way of searching for, understanding and removing any systematic problems.

CPU time-wise, the presented methods are limited by the noise
correlation matrix inversion, which requires \order{n_{pix}^3}
operation. 
This is the price to be paid if no assumption is made about possible
symmetries present in the map and a pixel-pixel noise correlation
matrix is required.
This obstacle may not be insurmountable, even without sacrificing 
the generality of the approach.
 In most of the
realistic situations such a matrix is expected to be rather sparse. Moreover,
from the knowledge of a scanning strategy and
characteristic correlation length in the time domain, it can be guessed
{\it a priori} which matrix elements need to be computed. 
We leave both issues for further investigation, noting only here that
extensive use of a supercomputer and the MADCAP package facilitates
computations of the maps containing up to $\sim 40,000$~\cite{Lee2001}.

\section{Acknowledgments}

RS and SH acknowledge support of NASA Grant NAG5-3941.
AHJ and JHPW acknowledge support from 
NASA LTSA Grant no.\ NAG5-6552 and NSF KDI Grant no.\ 9872979. 
PGF acknowledges support from the RS. 
BR and CDW acknowledge support from NASA GSRP
Grants no.\ S00-GSRP-032 and S00-GSRP-031.
RS also acknowledges help (in Poland) of Polish
State Committee for Scientific Research grant no.\ 2P03D01719.
Computing resources were provided by the
National Energy Research Computing Center at Lawrence Berkeley
National Laboratory.
MAXIMA is supported by NASA Grant NAG5-4454 and by the NSF through 
the Center for Particle
Astrophysics at UC Berkeley, NSF cooperative agreement AST-9120005.

\end{document}